\pdfoutput=1
\documentclass[10pt, conference, compsocconf]{IEEEtran}
\usepackage{float}

\usepackage{cite}

\ifCLASSINFOpdf
  \usepackage[pdftex]{graphicx}
\else
  \usepackage[dvips]{graphicx}
\fi
\usepackage[cmex10]{amsmath}
\ifCLASSOPTIONcompsoc
  \usepackage[caption=false,font=normalsize,labelfont=sf,textfont=sf]{subfig}
\else
  \usepackage[caption=false,font=footnotesize]{subfig}
\fi
\usepackage{fixltx2e}

\usepackage{amsfonts}

\makeatletter

\newcommand{\Rmnum}[1]{\expandafter\@slowromancap\romannumeral #1@}
\makeatother

\begin{document}
%
\title{Text Network Exploration via\\ Heterogeneous Web of Topics}




%
\author{\IEEEauthorblockN{Junxian He\IEEEauthorrefmark{1},
Ying Huang\IEEEauthorrefmark{2},
Changfeng Liu\IEEEauthorrefmark{1},
Jiaming Shen\IEEEauthorrefmark{3}, 
Yuting Jia\IEEEauthorrefmark{3}, 
Xinbing Wang\IEEEauthorrefmark{1}}
\IEEEauthorblockA{\IEEEauthorrefmark{1}Department of Electronic Engineering, Shanghai Jiao Tong University, Shanghai, China}
\IEEEauthorblockA{\IEEEauthorrefmark{2}Department of Automation, Shanghai Jiao Tong University, Shanghai, China}
\IEEEauthorblockA{\IEEEauthorrefmark{3}Department of Computer Science and Engineering, Shanghai Jiao Tong University, Shanghai, China\\
\{junxian831, hy941001, stevenevets, sjm940622, hnxxjyt, xwang8\}@sjtu.edu.cn
}}


\maketitle

\begin{abstract}
A text network refers to a data type that each vertex is associated with a text document and the relationship between documents is represented by edges. The proliferation of text networks such as hyperlinked webpages and academic citation networks has led to an increasing demand for quickly developing a general sense of a new text network, namely \emph{text network exploration}. In this paper, we address the problem of text network exploration through constructing a heterogeneous web of topics, which allows people to investigate a text network associating word level with document level. To achieve this, a probabilistic generative model for text and links is proposed, where three different relationships in the heterogeneous topic web are quantified.
We also develop a prototype demo system named \emph{TopicAtlas} to exhibit such heterogeneous topic web, and demonstrate how this system can facilitate the task of text network exploration. Extensive qualitative analyses are included to verify the effectiveness of this heterogeneous topic web. Besides, we validate our model on real-life text networks, showing that it preserves good performance on objective evaluation metrics. 

\end{abstract}

%
\IEEEpeerreviewmaketitle

\section{Introduction}
\begin{figure*}[!t]
\centering
\subfloat[Input text network]{\includegraphics[scale=0.16]{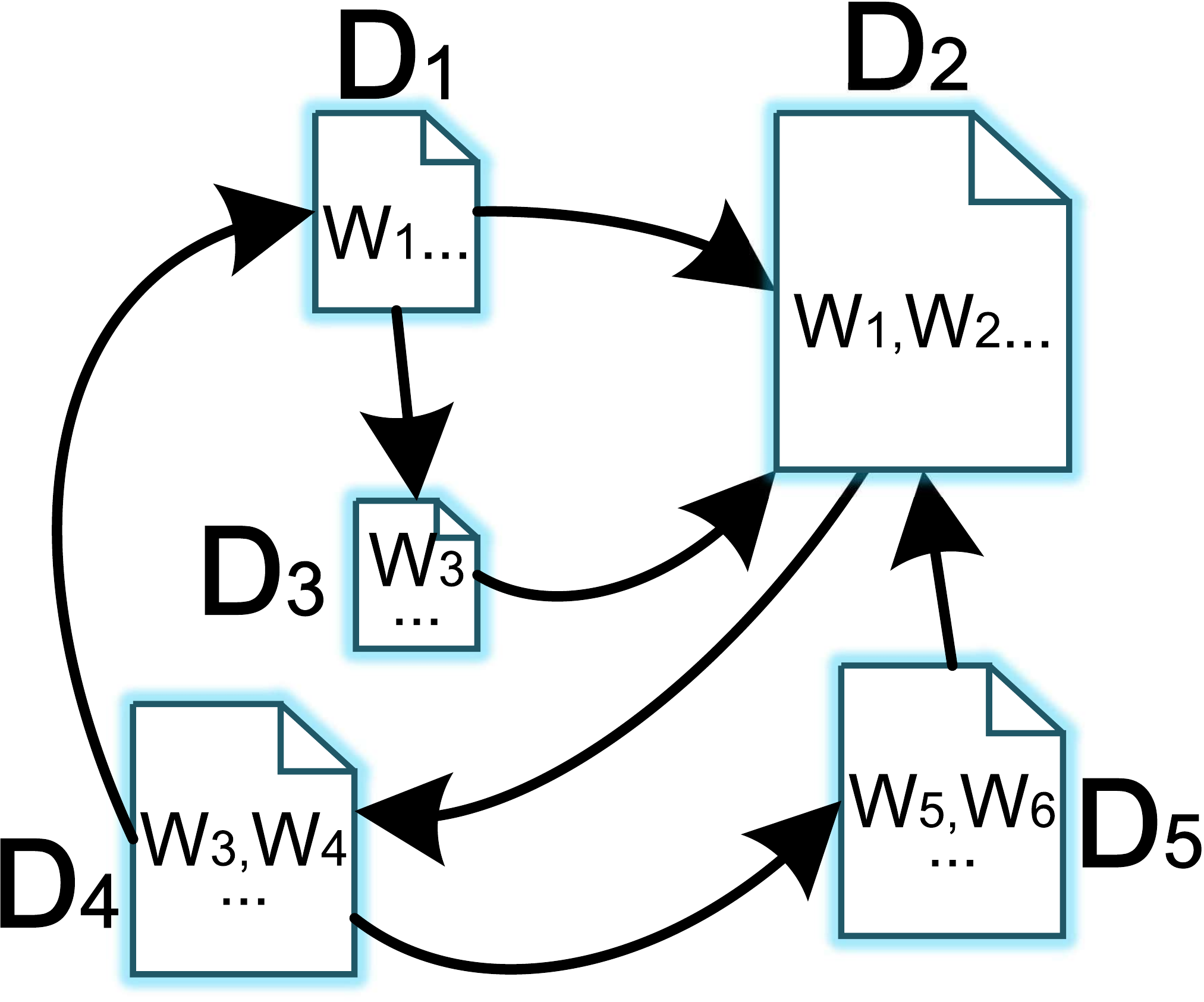}
\label{textnetwork}}
\hfil
\subfloat[Two parts of a document]{\includegraphics[scale=0.16]{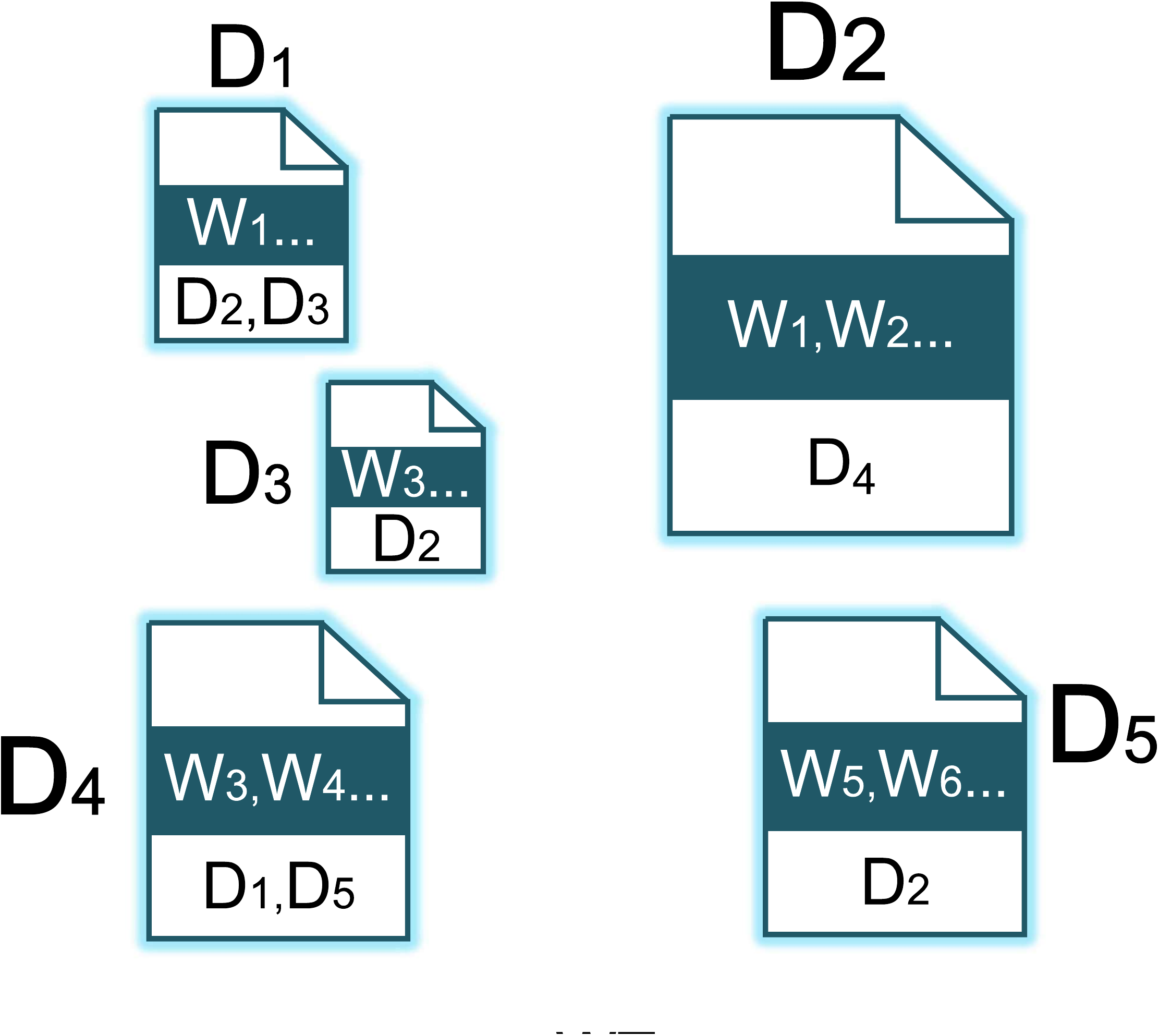}
\label{view}}
\hfil
\subfloat[WordTopic and DocTopic]{\includegraphics[scale=0.16]{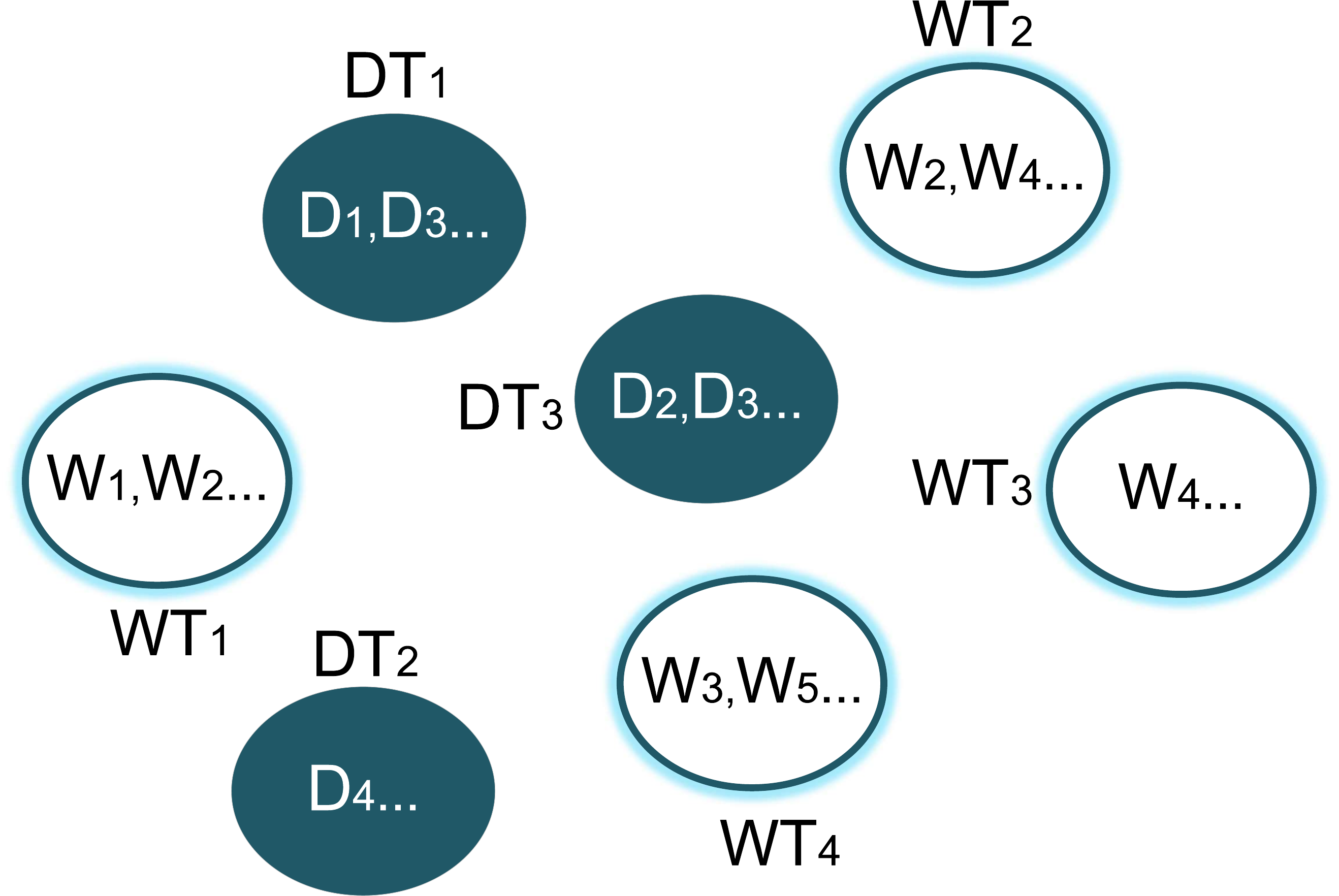}
\label{extendedtopic}}
\hfil
\subfloat[Heterogeneous topic web]{\includegraphics[scale=0.16]{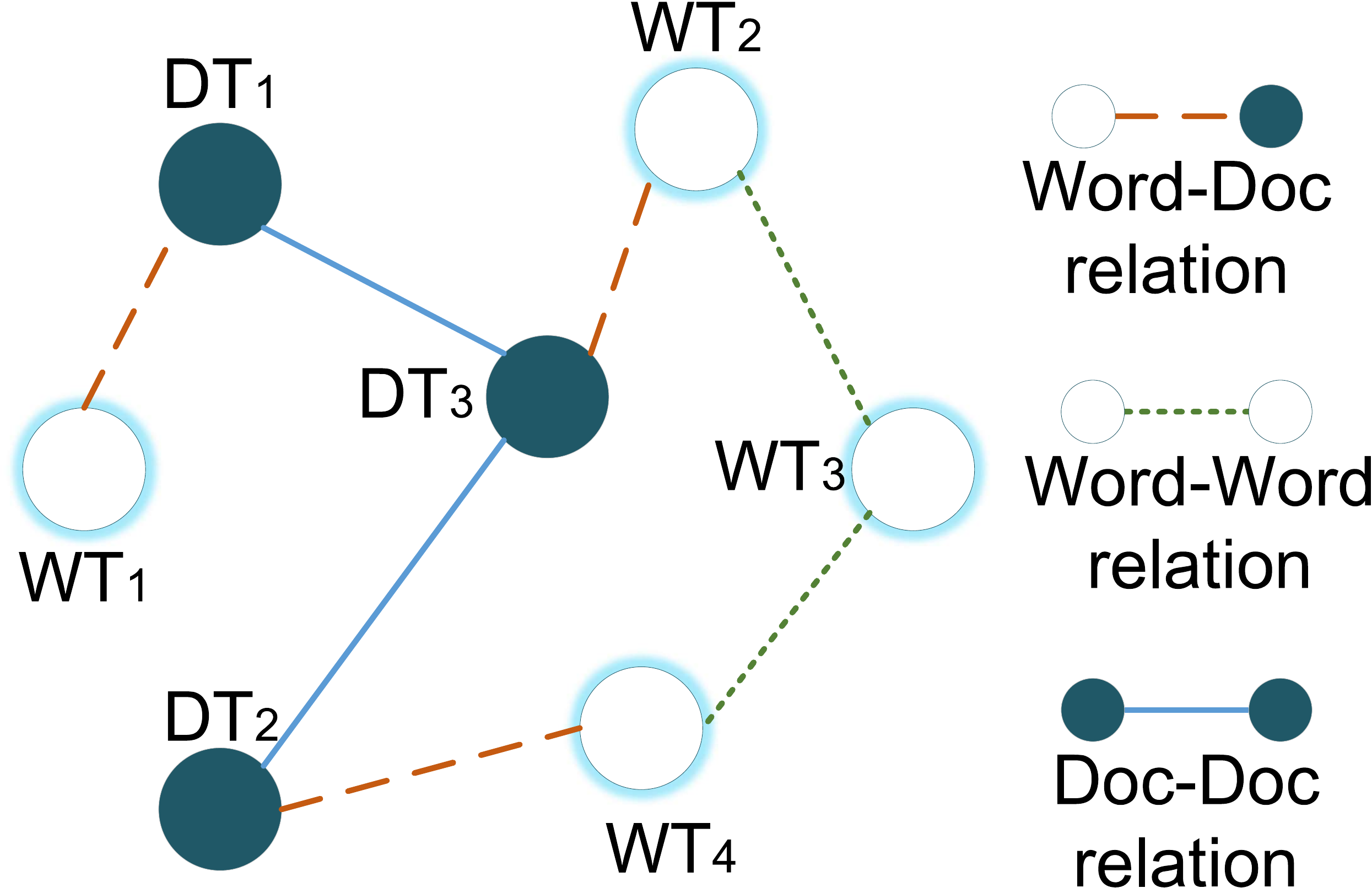}
\label{atlasDescrp}}
\caption{Illustration of some concepts. (a) Input text network. (b) Two parts of a document. W represents the ``word token'' part, and D below W represents ``document token'' part. (c) WordTopic (WT) and DocTopic (DT). (d) Heterogeneous topic web with two types of topics and three types of relationships.} \label{description} 
\end{figure*}
The information age has witnessed an increasing amount of unstructured data, most of which are in the form of text and possess high degrees of connectivity among themselves. We refer to this type of data as \emph{text network} as shown in Figure  \ref{textnetwork}. Such text networks are ubiquitous in the real world. Typical representatives include hyperlinked webpages, online social network with user profiles, and academic citation network. 

With the rapid increase of available text networks, the demand for exploring them quickly has continued to grow. When faced with a new or unfamiliar text network, people may first ask a basic question: ``What is there?". To answer this question, we resort to the notion of \emph{exploratory search} \cite{Marchionini2006} which is proposed to help people develop a general sense of the properties of new text network before embarking on more specific inquiries \cite{klein2015exploratory}. 

Due to its importance, exploratory search has been investigated intensively. For example, Sinclair et al. use word frequency lists, frequency distribution plots and keyword-in-context models to enhance computer-assisted reading \cite{sinclair2003computer}. More recently, a computational technique named ``topic modeling'' achieves great success through providing insight into a corpus' contents~\cite{gretarsson2012topicnets,alexander2014serendip,klein2015exploratory}. Nevertheless, existing topic models are still far from adequate for text network exploration since the significance of topics represented only by words is limited for exploration task without an insight on document level.

To address this problem, we view each document as a ``bag of links'' \cite{linklda,citationlda}. For example, in the academic paper network, a paper with $k$ references is viewed as a document with $k$ ``link tokens''. Then, we can model these documents within a topic model framework where a new type of ``topics'' characterized by distributions over documents arises and important documents are assigned with high probabilities. By combining ``word token'' and ``document token'', each document is composed of two parts as shown in Figure \ref{view}, and two different types of topics are included as illustrated in Figure \ref{extendedtopic}. To distinguish the two categories of topics, we call them \emph{WordTopic} and \emph{DocTopic} respectively.

However, it is still inconvenient to explore a text network since users can only inspect the individual topic in isolation. Therefore, we expect to uncover the relations between topics to enable users to examine not only a topic itself but also the related fields and important documents. With that in mind, a complete heterogeneous topic web which displays three different types of relationships as described in Figure \ref{atlasDescrp} is indispensable. Although the relationship between WordTopics (\emph{Word-Word relation}) has been investigated previously~\cite{blei2006correlated,pairwise,chang2009relational,ITM,nallapati2011topicflow,citationlda,weng2014topic,wang2015constructing}, the connections between DocTopic and DocTopic (\emph{Doc-Doc relation}) and WordTopic and DocTopic (\emph{Word-Doc relation}) have not been studied before.

To construct such heterogeneous topic web, we propose a probabilistic generative model called \textbf{MHT} (Model for Heterogeneous Topic web), where all three relationships are quantified. Our experiments on two academic citation networks demonstrate that MHT not only produces reliable heterogeneous topic web with high-quality topics but also possesses strong generalizability and predictive power. 

Furthermore, we build \textbf{TopicAtlas}, a prototype demo system for convenient navigation in heterogeneous topic web. TopicAtlas displays Word-Word relation, Doc-Doc relation, and Word-Doc relation in a unified framework. With TopicAtlas, users are able to freely wander around the text network via WordTopics and DocTopics.

To summarize, our contributions are three folds: 
\begin{itemize}
  \item To the best of our knowledge, we are the first to present the idea of heterogeneous web of topics and construct it successfully. 
  \item We propose MHT, a probabilistic generative model that helps extract two types of topics along with their heterogeneous relationships.   
  \item We develop TopicAtlas, a prototype system for text network exploration. TopicAtlas allows users to investigate the heterogeneous topic web with details and explore text network easily.
\end{itemize}

The rest of paper is organized as follows. In section \ref{related work}, we discuss some related works. We introduce MHT and its inference in section \ref{secMHT} and \ref{secLearning}. In section \ref{secExperiment}, we conduct the experiment and evaluate our model. Finally, we summarize this paper and discuss some future works in section \ref{secConclusion}. 

\section{Related Work}
\noindent \textbf{In terms of exploratory search}. When dealing with large collections of digitized historical documents, very often only little is known about the quantity, coverage and relations of its content. In order to get an overview, exploring the data beyond simple ``lookup'' approaches is needed. The notion of exploratory search has been introduced to cover such cases~\cite{Marchionini2006}. 

Chaney and Blei~\cite{chaney2012visualizing} make an early effort in exploratory search via visualizing traditional topic models, where a navigator of documents is created and allows users to explore the hidden structure. Gretarsson et al. build a relatively mature system called TopicNets \cite{gretarsson2012topicnets}, which enables users to visualize individual document sections and their relations within the global topic document. Maiya et al.~\cite{maiya2014topic} build the topic similarity network for exploration and recognize how topics form large themes. Recently, Jahnichen et al. develop a complete framework in this field~\cite{exploratory_jahnichen}, they depict probability distributions as tag clouds and permit the identification of related topic groups or outliers. 

While the works mentioned above convey some information visually, these approaches consider the data as isolated-document corpus rather than linked text networks. With only text they cannot conduct a serious analysis for a text network on a document level. Specifically, although some of them are able to retrieve \textit{topic-related} documents, there is no possibility for them to identify \textit{topic-significant} documents, which are more crucial in exploratory search. Therefore, we introduce DocTopic and propose the idea of heterogeneous topic web to enable users to keep track of related topic groups, relevant documents and significant documents.

\label{related work}
\noindent \textbf{In terms of topic modeling}. Topic models are proposed to address the problem of topic identification in large document collections. In topic models, each document is associated with a topic distribution and each topic is associated with a word distribution. Two popular models in this field are Probabilistic Latent Semantic Analysis (PLSA) \cite{plsa} and Latent Dirichlet Allocation (LDA) \cite{blei2003latent}. They are both
generative and unsupervised models, introducing latent topics into the generative process.

However, traditional topic models only consider text and ignore the significant link information. Recently, some variants of topic models are proposed for jointly analyzing text and links. A major part of them models the link information as evidence of content similarity between two documents~\cite{ITM2,mei2008topic,linkplsalda,chang2009relational,ITM,liu2009topic,nallapati2011topicflow,le2014probabilistic}, but this kind of approach is not able to detect important documents with respect to a specific topic. Another categories of methods which generate the links from DocTopics can recognize significant documents~\cite{phits2,phits,linklda,pairwise,citationlda}. But these works fail to construct a complete heterogeneous topic web composed of WordTopic, DocTopic and three different types of relations among them. Although the connection between WordTopics has been investigated before~\cite{blei2006correlated,pairwise,chang2009relational,ITM,nallapati2011topicflow,citationlda,weng2014topic,wang2015constructing}, to the best of our knowledge, we are the first to model two types of topics and three types of relations jointly and build the heterogeneous topic web successfully.

\section{Model for Heterogeneous Topic Web}
\label{secMHT}

\begin{figure}[!t]
\begin{center}
\includegraphics[scale=0.5]{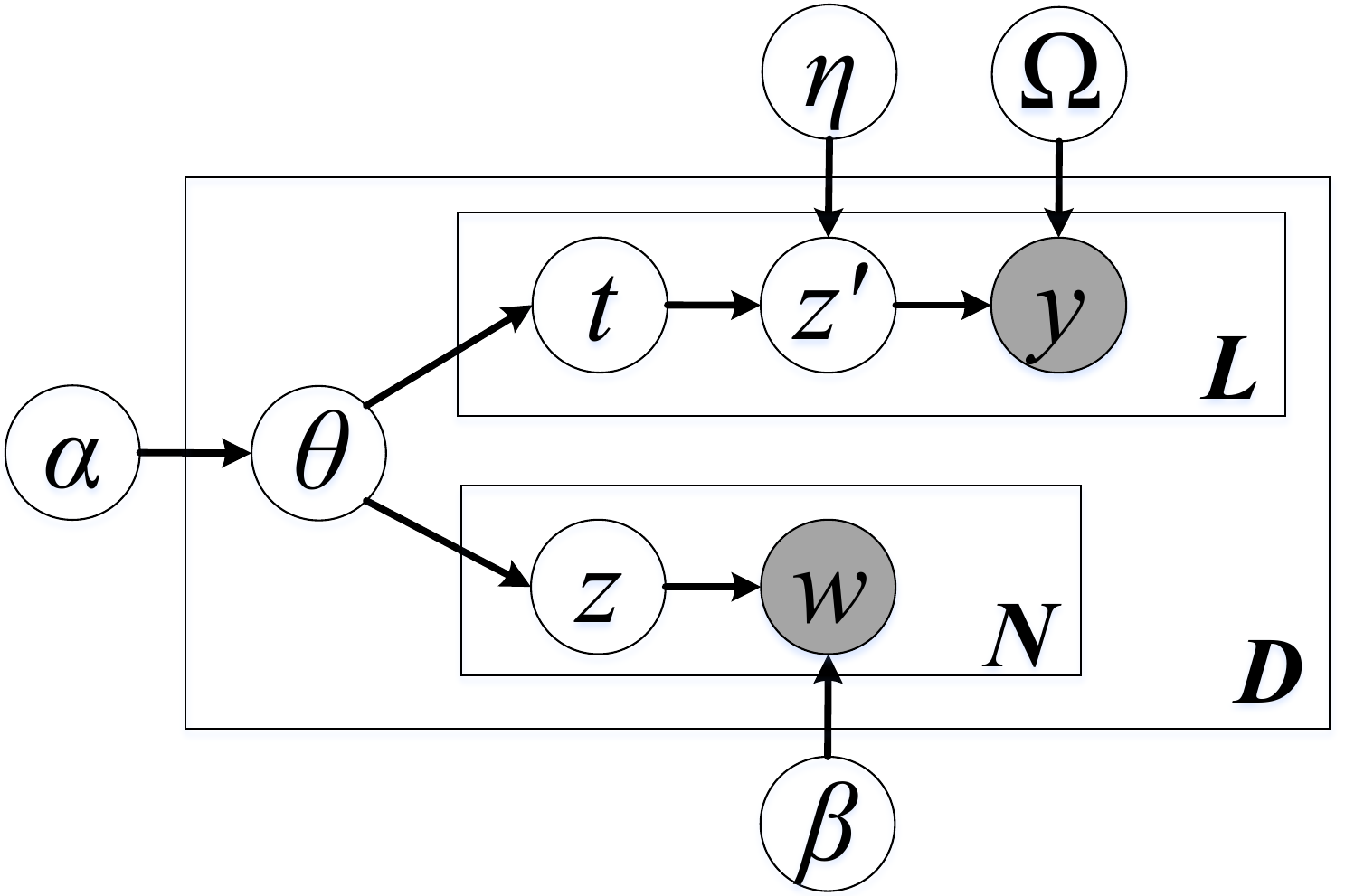}
\caption{Graphical Representation of MHT}
\label{model}
\end{center}
\end{figure}

In this part we describe the framework and generative process of MHT (Model for Heterogeneous Topic web), whose graphical representation is illustrated in Figure \ref{model}. 

\subsection{Framework}
We consider the input text network as a graph $G(V,E)$, where $V$ is the set of document vertices and $E$ is the set of directed edges or links. $v_i\in V$ represents the $i^{th}$ document and $e_{ij}\in E$ connects two vertices $v_i$ and $v_j$. Each document is associated with a bag of words and a bag of links. We denote $w_{in}$ as the $n^{th}$ word token in document $v_i$, and $y_{il}$ expresses the $l^{th}$ link token (document token) in $v_i$.

In classical topic models each document is associated with a document specific topic distribution, which is used to draw a topic for each word in the generative process. Note that the ``topic'' here actually represents WordTopic in our notation framework. Similarly, adopt the assumption of ``bag of links'' and each document is associated with a DocTopic distribution, which can generate linked documents. Since these two distributions are totally different, some transition procedure between them is required to jointly model text and links.

Based on the discussion above, we employ a transition distribution over DocTopics $\eta$ to depict the relation between the two types of topics. 

\subsection{Generative Process}
Details for full generative process of our proposed model MHT are demonstrated below.\\
\\
For each document $v_i$, where $i=1,\cdots, D$:
\begin{enumerate}
    \item Generate WordTopic distribution: 
    $\theta_i\sim Dir(\cdot | \alpha)$
    \item For each word $w_{in}$, where $n=1, \cdots, N_i$:
        \begin{enumerate}
        \item Draw a WordTopic:
        $z_{in}\sim Mult(\cdot | \theta_i)$
        \item Draw a word:
        $w_{in}\sim Mult(\cdot | \beta_{z_{in}})$
        \end{enumerate}
    \item For each link $y_{il}$, where $l=1, \cdots, L_i$:
        \begin{enumerate}
        \item Draw a transition topic:
        $t_{il}\sim Mult(\cdot | \theta_i)$
        \item Draw a DocTopic:
        $z^{\prime}_{il}\sim Mult(\cdot | \eta_{t_{il}})$
        \item Draw a linked document:
       $y_{il}\sim Mult(\cdot | \Omega_{z^{\prime}_{il}})$
        \end{enumerate}
\end{enumerate}

Step 1 and Step 2 are the same as classical topic model to generate words. A major distinction of MHT from other models is Step 3, where we employ a transition latent variable $t$ as an ``intermediary'' from WordTopic domain to DocTopic domain. In this transition stage, we introduce a transition parameter $\eta$ to express the relation between WordTopic and DocTopic so that the generation of DocTopic is equivalent to drawing it from $\theta\eta$. Thus $\eta$ serves as a transition matrix from $\theta$ to a ``spurious'' underlying mixed DocTopic distribution $\theta^{\prime}$. More specifically, for a given WordTopic $k$, the value of $\eta_{kk^{\prime}}$ indicates the probability for generating DocTopic $k^{\prime}$, i.e. $p(z^{\prime}=k^{\prime} | z=k)=\eta_{kk^{\prime}}$. With that in mind, we can see how $\eta$ works on transforming WordTopic domain into DocTopic domain. 

\section{Model Learning}
\label{secLearning}

To learn MHT, we resort to the variational EM inference method. For each document $v_{i}$, we use a fully factorized variational distribution to approximate the posterior distribution:
\begin{small}
\begin{equation}
\label{approx_posterior}
\begin{aligned}
q(\theta_i, z_i, t_i, z^{\prime}_i)&=q(\theta_i | \gamma_i)\prod^{N_i}_{n=1}q(z_{in} | \phi_{in})\\
&\times\prod^{L_i}_{l=1}q(t_{il}|\lambda_{il})\prod^{L_i}_{l=1}q(z^{\prime}_{il}|\sigma_{il}),
\end{aligned}
\end{equation}
\end{small}
where $q(\theta_i | \gamma_i)$ is Dirichlet distribution and $q(z_{in} | \phi_{in})$, $q(t_{il}|\lambda_{il})$ and $q(z^{\prime}_{il}|\sigma_{il})$ are all multinomial distributions.  Then we will try to maximize the evidence lower bound defined by:
\begin{small}
\begin{equation}
\label{elbo}
\begin{aligned}
\text{ELBO} =& \sum_{i=1}^D(\mathbb{E}_q[\log p(\theta_i, z_i, t_i, z^{\prime}_i,\mathbf{w_i, y_i} | \alpha,\eta,\beta,\Omega)]\\
&-\mathbb{E}_q[\log q(\theta_i, z_i, t_i, z^{\prime}_i)]),
\end{aligned}
\end{equation}
\end{small}

In the E-step, we update $\gamma, \phi, \lambda \text{ and } \sigma$ iteratively to approximate the posterior distribution. Then, in the M-step, $\alpha, \beta, \eta \text{ and } \Omega$ are renewed to maximize ELBO. Due to the limitation of space, we only provide crucial equations here.
\begin{small}
\begin{equation}
\label{phi}
\phi_{ink}\propto \beta_{kx}\text{exp}(\Psi(\gamma_{ik})).
\end{equation}
\end{small}
\begin{small}
\begin{equation}
\label{gamma}
\gamma_{ik}=\alpha_k+\sum^{N_i}_{n=1}\phi_{ink}+\sum^{L_i}_{l=1}\lambda_{ilk}.
\end{equation}
\end{small}
\begin{small}
\begin{equation}
\label{lambda}
\lambda_{ilk}\propto \text{exp}(\Psi(\gamma_{ik})+\sum^{K_y}_{k^{\prime}=1}\sigma_{ilk^{\prime}}\text{log}\eta_{kk^{\prime}}).
\end{equation}
\end{small}
\begin{small}
\begin{equation}
\label{sigma}
\sigma_{ilk^{\prime}}\propto \Omega_{k^{\prime}d}\text{exp}(\sum^{K_w}_{k=1}\lambda_{ilk}\text{log}\eta_{kk^{\prime}}).
\end{equation}
\end{small}
\begin{small}
\begin{equation}
\label{beta}
\beta_{kx}\propto \sum^D_{i=1}\sum^{N_i}_{n=1}w^{x}_{in}\phi_{ink}.
\end{equation}
\end{small}
\begin{small}
\begin{equation}
\label{eta}
\eta_{kk^{\prime}}\propto\sum^D_{i=1}\sum^{L_i}_{l=1}\sigma_{ilk^{\prime}}\lambda_{ilk}.
\end{equation}
\end{small}
\begin{small}
\begin{equation}
\label{omega}
\Omega_{k^{\prime}d}\propto \sum^D_{i=1}\sum^{L_i}_{l=1}y^{d}_{il}\sigma_{ilk^{\prime}}.
\end{equation}
\end{small}
Here, $\Psi(\cdot)$ is the digmma function, $w^x_{in}=1$ if $w_{in}=x$, and 0 otherwise. Likewise, $y^d_{il}=1$ if $y_{il}=d$, and 0 otherwise. $\alpha$ is updated by Newton-Raphson algorithm, the interested readers may refer to~\cite{blei2003latent}.

First, for each document, we execute step (\ref{phi}) to (\ref{sigma}) iteratively until convergence. And then we update $\alpha$, $\beta$, $\eta$ and $\Omega$.
The whole process is in an outer loop until the lower bound ELBO converges.

\section{Experiments}
\label{secExperiment}
In this section, we demonstrate how our proposed system -- TopicAtlas effectively explores text networks. For repeatability, the codes, datasets, results and the demo TopicAtlas are available to the public\footnote{https://river459.github.io/research/}.

\subsection{Datasets}
We use the following two datasets in our experiments: 

\textbf{ACL Anthology Network (AAN).} AAN~\cite{aan_source} is a public scientific literature dataset in the Natural Language Processing (NLP) field with $20,989$ abstracts of papers and $125,934$ citations. 

%
\textbf{CiteseerX.} CiteseerX\footnote{http://citeseer.ist.psu.edu/oai.html} is a well-known scientific literature digital library that primarily focuses on the literature in computer and information science. We collect a subset of CiteseerX dataset, which includes the abstracts of $716,800$ documents and $1,760,574$ links. 

 \subsection{Parameter Setting}
On the task of exploring heterogeneous topic web, we first need to select a reasonable topic number, which is a non-trivial task in topic models. To achieve this, we first preprocess the data using classical LDA model with varying topic numbers and evaluate the topic interpretability in terms of the topic coherence score \cite{mimno2011optimizing}. Among the candidate topic numbers 50, 70, 90, 110, 130, and 150, topic number 70 leads to the highest topic coherence score for both AAN and CiteseerX. For simplicity, we set the topic number of WordTopic and DocTopic equal. Therefore, we implement MHT with 70 WordTopics and 70 DocTopics to explore the text networks in the two datasets. In addition, we follow the convention of \cite{griffiths2004finding} and initialize $\alpha = 0.01$. The parameters $\eta$, $\beta$ and $\Omega$ are randomly initialized since we do not have any prior knowledge. 

 Furthermore, as discussed above, we use variational EM inference to learn the parameters in MHT. In our experiments, for both datasets the inner variational inference loop terminates when the fractional increase of ELBO is less than $10^{-9}$ in two successive iterations, or the number of iterations exceeds 100. For the outer EM loop, we stop it when the relative increment ratio is less than $10^{-4}$, or the number of iterations exceeds 50. 


\subsection{Heterogeneous Topic Web Construction}
We use co-occurrence probability to quantify the strength of the three types of relations in heterogeneous topic web.

\textbf{Word-Word Relation Strength.} Since we assume the generation of WordTopics is independent with each other when the document $v$ is given, the Word-Word relation strength can be calculated as follows: 
\begin{small}
\begin{equation}
\begin{aligned}
p(z_1=k_1, z_2=k_2 | D)&=\sum_{z^{\prime}}\sum_ip(z^{\prime} | D) p(v_i | z^{\prime}; D)\\
&\times p(z_1=k_1 | v_i; D)\\
&\times p(z_2=k_2 | v_i; D),
\end{aligned}
\end{equation}
\end{small}
where $p(z | v; D)$ and $p(v | z^{\prime}; D)$ can be obtained from $\theta$ and $\Omega$ respectively. Posterior expectation of $\theta$ is given by:
\begin{small}
\begin{equation}
\label{theta1}
\theta_{ik}=\frac{\#( v=i, z=k) +\alpha_k}{\sum^{K_w}_{k^*=1}(\#( v=i, z=k^*)+\alpha_{k^*})},
\end{equation}
\end{small}
where $\#( v=i, z=k)$ represents the number of words assigned with WordTopic $k$ in document $v_i$ and the assignment can be obtained from $\phi$. $K_w$ is the number of WordTopics.

In addition, the empirical posterior distribution over DocTopics can be computed as:
\begin{small}
\begin{equation}
p(z^{\prime}=k^{\prime} | D)=\frac{\#(z^{\prime}=k^{\prime})}{\sum_{k^{*}}\#(z^{\prime}=k^{*})},
\end{equation}
\end{small}
where $\#(z^{\prime}=k^{\prime})$ represents the number of documents assigned with DocTopic $k^{\prime}$ and can be obtained from $\sigma$.

\textbf{Doc-Doc Relation Strength.} Based on the assumption that DocTopics are generated independently given a WordTopic, we can compute Doc-Doc relation strength as:
\begin{small}
\begin{equation}
\begin{aligned}
p(z^{\prime}_1=k^{\prime}_1, z^{\prime}_2=k^{\prime}_2 | D) &=\sum_zp(z | D)p(z^{\prime}_1=k^{\prime}_1 | z; D)\\
& \times p(z^{\prime}_2=k^{\prime}_2 | z; D).
\end{aligned}
\end{equation}
\end{small}

$\eta$ represents $p(z^{\prime} | z; D)$ and similarly the empirical posterior distribution over WordTopics is given by:
\begin{small}
\begin{equation}
p(z=k | D)=\frac{\#(z=k)}{\sum_{k^{*}}\#(z=k^{*})}.
\end{equation}
\end{small}

\textbf{Word-Doc Relation Strength.} Word-Doc relation strength can be easily computed by Bayes' theorem:
\begin{small}
\begin{equation}
p(z=k, z^{\prime}=k^{\prime} | D)=p(z^{\prime}=k^{\prime} | z=k; D) p(z=k | D).
\end{equation}
\end{small}

\textbf{Summarizing DocTopic.}
While top words are able to represent WordTopic explicitly, on the document side there are only distributions over documents to express DocTopics. However, generally it would be preferable to summarize topics with a few words~\cite{chang2009reading}. With that in mind, we leverage the words in abstracts to summarize DocTopics. Specifically, for a given DocTopic $k^{\prime}$, we compute the expectancy of word $w$ as:
\begin{small}
\begin{equation}
\setlength{\abovedisplayskip}{0pt}
\setlength{\belowdisplayskip}{0pt}
\label{ }
\mathbb{E}(w|z^{\prime}={k^{\prime}})=\sum^D_{d=1}\Omega_{k^{\prime}d}\cdot\#(w,d).
\end{equation}
\end{small}
Then the words with high expectancy are selected as \emph{indicative words} of this DocTopic, which will be displayed in our demo system TopicAtlas.

\subsection{TopicAtlas}
\begin{figure}[!hbtp]
\begin{center}
\includegraphics[scale=0.15]{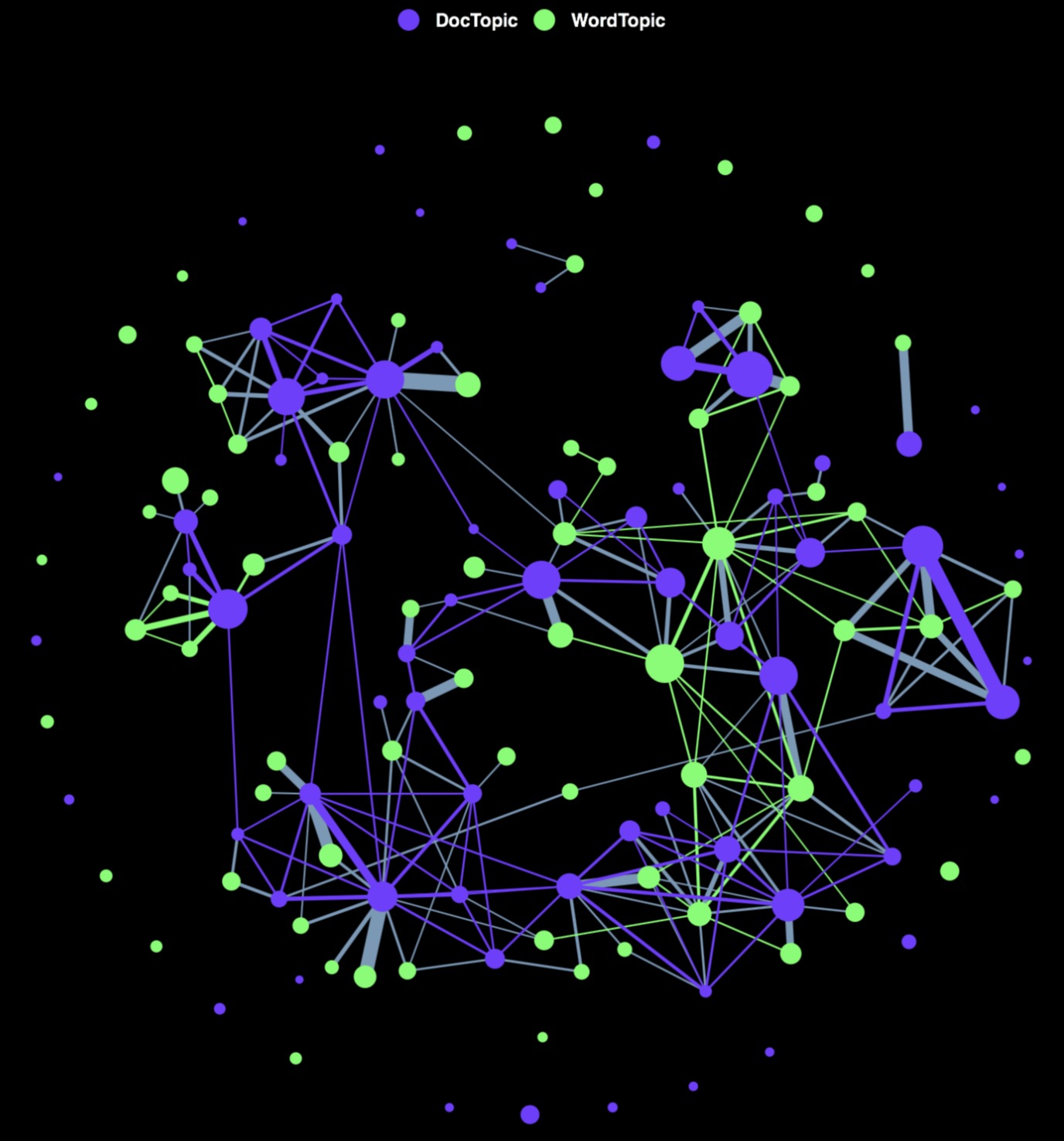}
\caption{An overview of TopicAtlas. Different colors indicate different types of topics, and the node size expresses the dominance of corresponding topic. Thickness of edges is proportionate to relation strength (best seen in color).}
\label{overview}
\end{center}
\end{figure}

\begin{figure*}[!h]
\centering
\subfloat[Word-Word subgraph]{\includegraphics[scale=0.25]{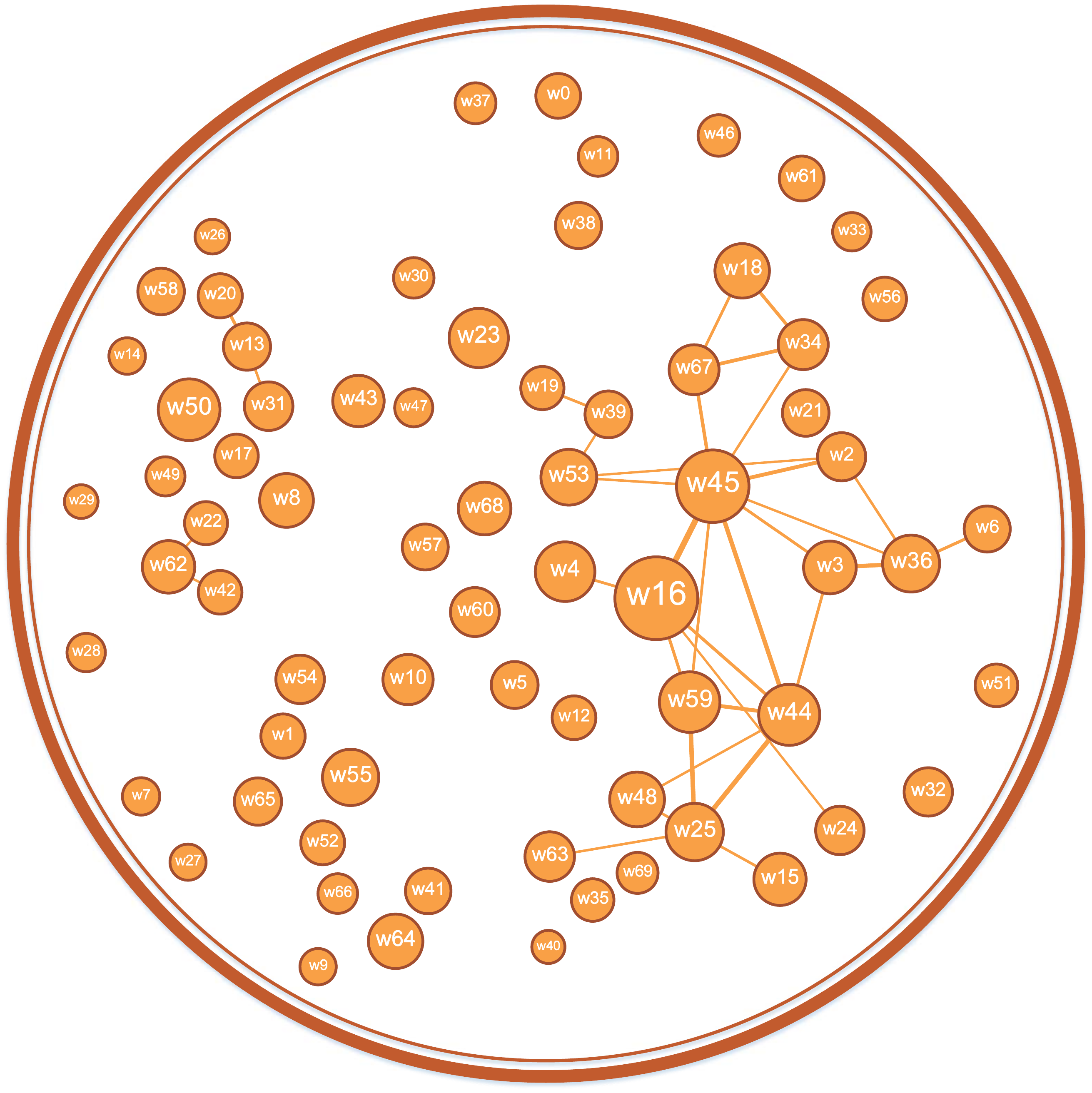}
\label{word-word graph}}
\hfil
\subfloat[Doc-Doc subgraph]{\includegraphics[scale=0.25]{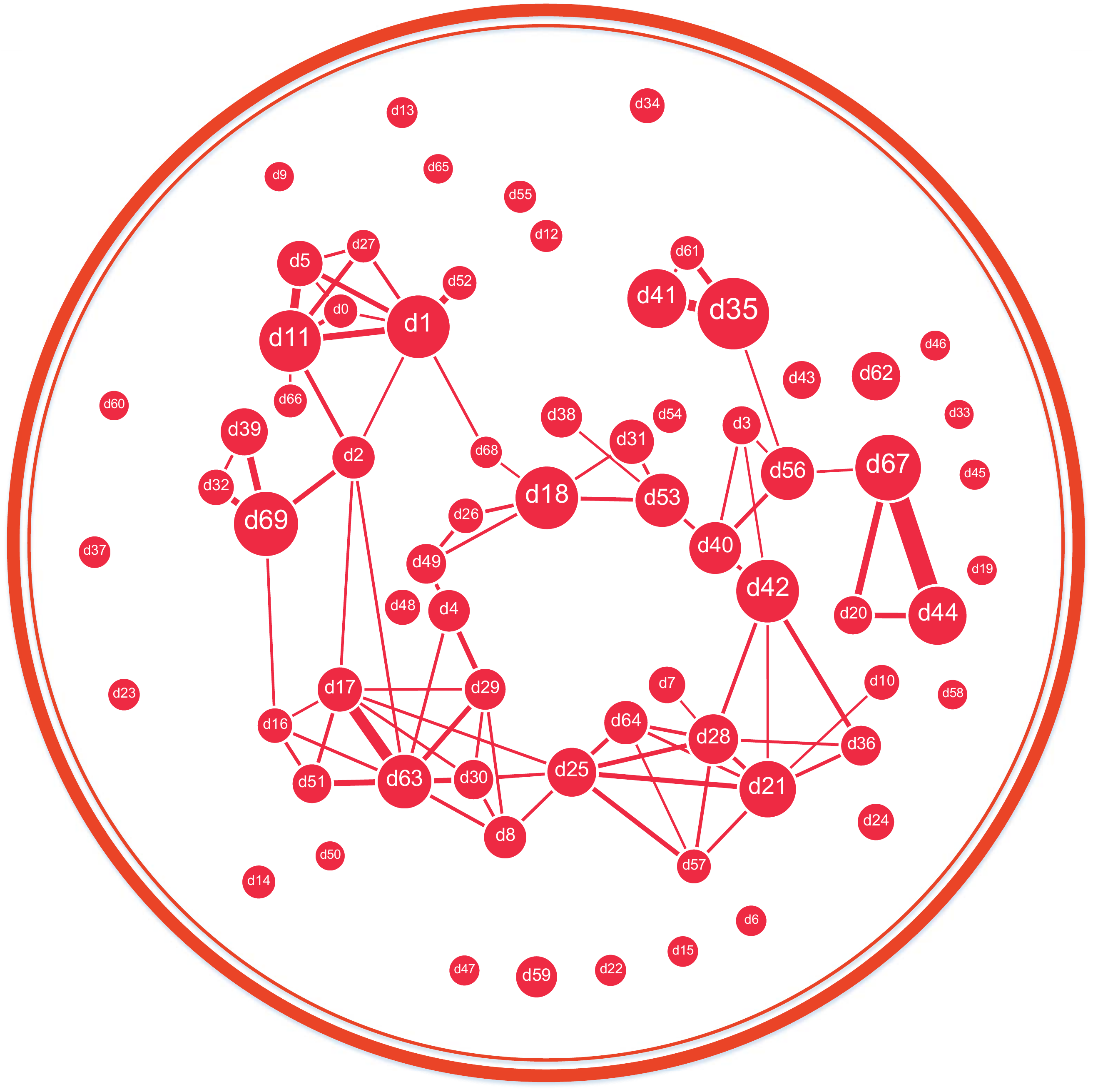}
\label{doc-doc graph}}
\caption{Subgraphs of heterogeneous topic web: (a) Word-Word Subgraph and (b) Doc-Doc Subgraph (best seen in color). The yellow nodes in the Word-Word subgraph represent WordTopics and the red nodes in the Doc-Doc subgraph represent DocTopics} \label{subgraph}
\end{figure*}
We design TopicAtlas based on the constructed heterogeneous topic web. An overview of TopicAtlas is displayed in Figure \ref{overview}. Aiming to help users navigate in an unfamiliar text network, TopicAtlas has the following features: 

\textbf{Topic Landscape Exhibition.} We display top 10 keywords for each WordTopic, and top 5 representative documents and top 10 indicative words for each DocTopic. The diameters of topic vertices express their corresponding \emph{topic dominance} or \emph{topic importance}, which is indicated by $p(z | D)$ for each WordTopic and $p(z^{\prime} | D)$ for each DocTopic. 

\textbf{Accurate Relationship.} The three types of relations correspond to three types of edges in the graph. The weights of these edges are the ratio of the co-occurrence probability we calculate to the prior probability of a random edge (0.0002). The thickness of the edges is proportionate to these values and we remove those whose weights are negligible. 

\subsection{Text Network Exploration via Heterogeneous Topic Web}
In this part, we engage in an in-depth exploration of the heterogeneous topic web. To facilitate the analytic reasoning, three auxiliary subgraphs of TopicAtlas are presented here: \emph{Word-Word} subgraph, \emph{Doc-Doc} subgraph and \emph{Word-Doc} subgraph. As the name suggests, Word-Word subgraph only includes the edges between WordTopics, Doc-Doc subgraph contains merely the edges between DocTopics, and Word-Doc subgraph displays edges between WordTopics and DocTopics. Due to the limitation of the space, we only give analysis for CiteseerX here and interested readers can refer to the public demo for the AAN TopicAtlas.


\begin{figure}[!t]
\begin{center}
\includegraphics[scale=0.38]{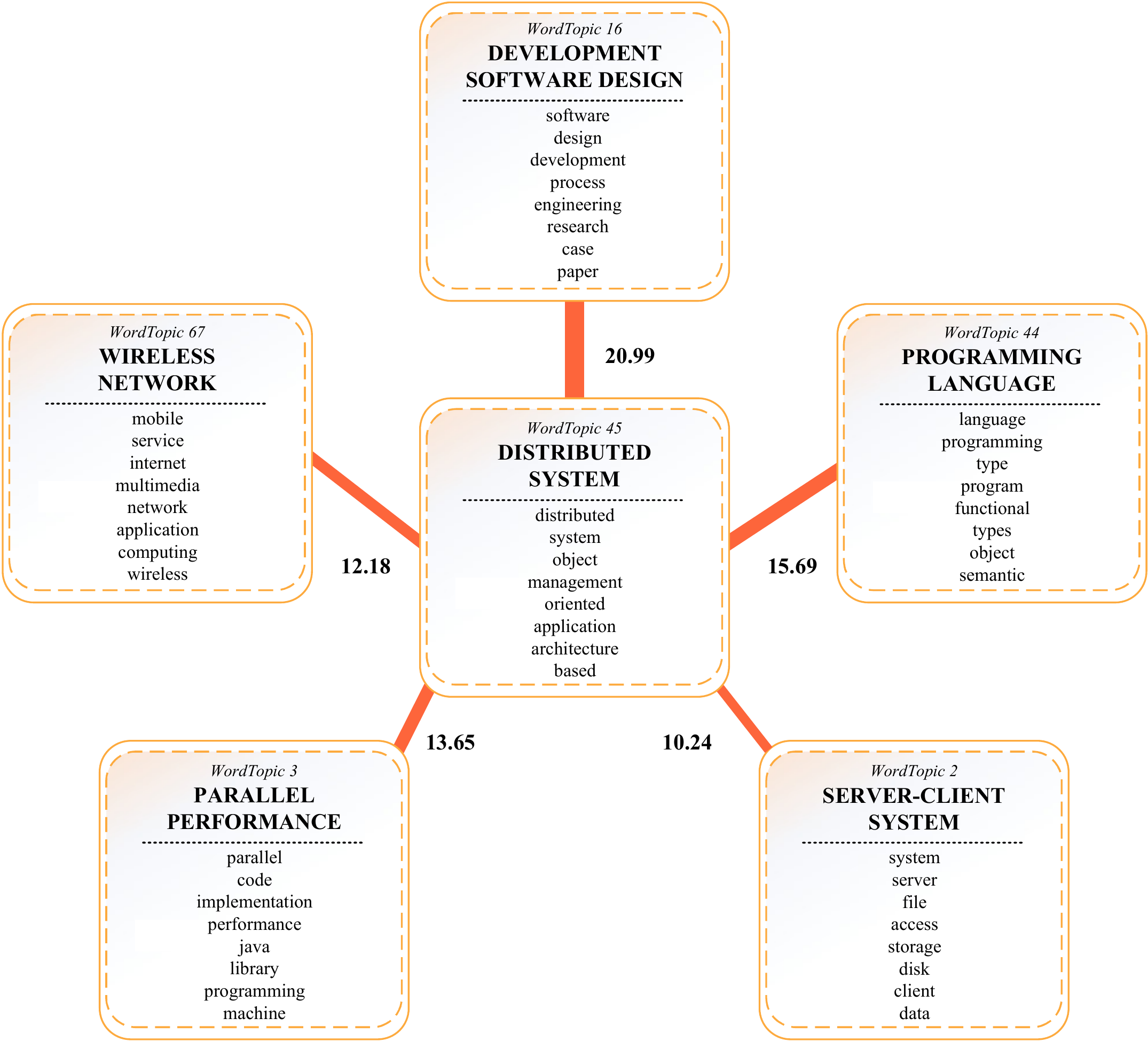}
\caption{``\emph{Distributed system}'' example. These topics are labeled manually.}
\label{word-word relation}
\end{center}
\end{figure}
\subsubsection{\textbf{Word-Word Relation}} As shown in Figure \ref{word-word graph}, 62.87\% of WordTopic nodes have no connection with other WordTopic nodes, which implies that one paper mainly focuses on one WordTopic. This phenomenon agrees with our intuition: most of high quality scientific papers show clear themes.

Though the connection between WordTopics is not strong, there are still a few nodes which link to multiple WordTopics worth investigating. On the basis of previous recognition that the content of documents is generally ``pure'', we believe that those WordTopics which enjoy high co-occurrence probability with various other WordTopics are foundation of certain scientific fields. In Figure \ref{word-word graph}, WordTopic w45 (degree: 9), w44(degree: 6), w16 (degree: 5), and w25 (degree: 5) have the highest degrees. The corresponding WordTopics are ``\emph{distributed system}'', ``\emph{programming language}'' , ``\emph{software design}'', and ``\emph{semantic reasoning}''. Obviously they are all general and basic. Take ``\emph{distributed system}'' as an example, distributed system achieves efficiency improvement of solving computational problems and therefore has broad applications in different fields such as telephone networks, routing algorithms, network file system, etc. As a case study, we show WordTopic w45 ``\emph{distributed system}'' and its related WordTopics in Figure \ref{word-word relation}.

\subsubsection{\textbf{Doc-Doc Relation}} The DocTopics are closely connected as shown in the Figure \ref{doc-doc graph}, which indicates that authors tend to cover multiple DocTopics in the reference list. It is intuitive because a comprehensive reference section is desired for most authors.
Furthermore, since ubiquitous techniques are likely to be cited in a variety of distinct domains, we expect nodes with high degrees in the Doc-Doc subgraph represent DocTopics about universal principle and method. In  Figure~\ref{doc-doc graph}, the top four highest-degree nodes are DocTopic d63 (degree: 11), d28 (degree:7), d21 (degree:7), d17 (degree:7) and they represent ``\emph{linear system method}'', ``\emph{logic programming}'', ``\emph{model checking}'' and ``\emph{conservation law}'' respectively. Unsurprisingly, these DocTopics are basic techniques and laws.



\begin{figure}[t]
\begin{center}
\includegraphics[scale=0.18]{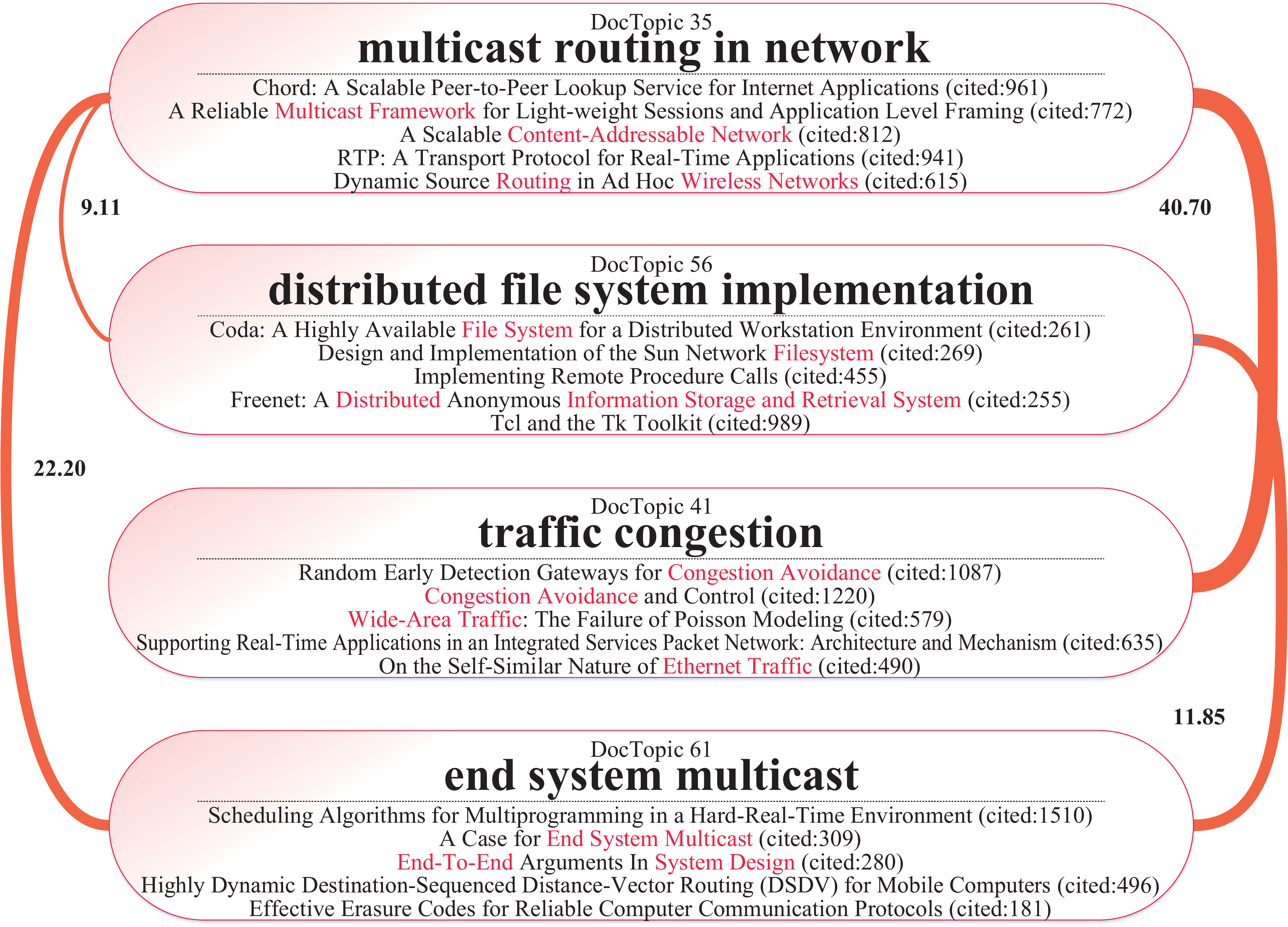}
\caption{``Multicast routing in network'' example. These topics are labeled manually. For each document, we display its citation number in our dataset.}
\label{doc doc relation}
\end{center}
\end{figure}

In addition to examining DocTopics from a global perspective, inspecting details of specific DocTopic provides insight into a text network on the document level. The DocTopic allows us to assess topic-aware impact of papers since the top documents in one given DocTopic are generally the most popular and representative ones. 
In  Figure~\ref{doc doc relation} we list top 5 documents in the most dominant DocTopic d35 and its neighbours d41, d56, d61. 

\begin{figure*}[!h]
\begin{center}
\includegraphics[scale=0.3]{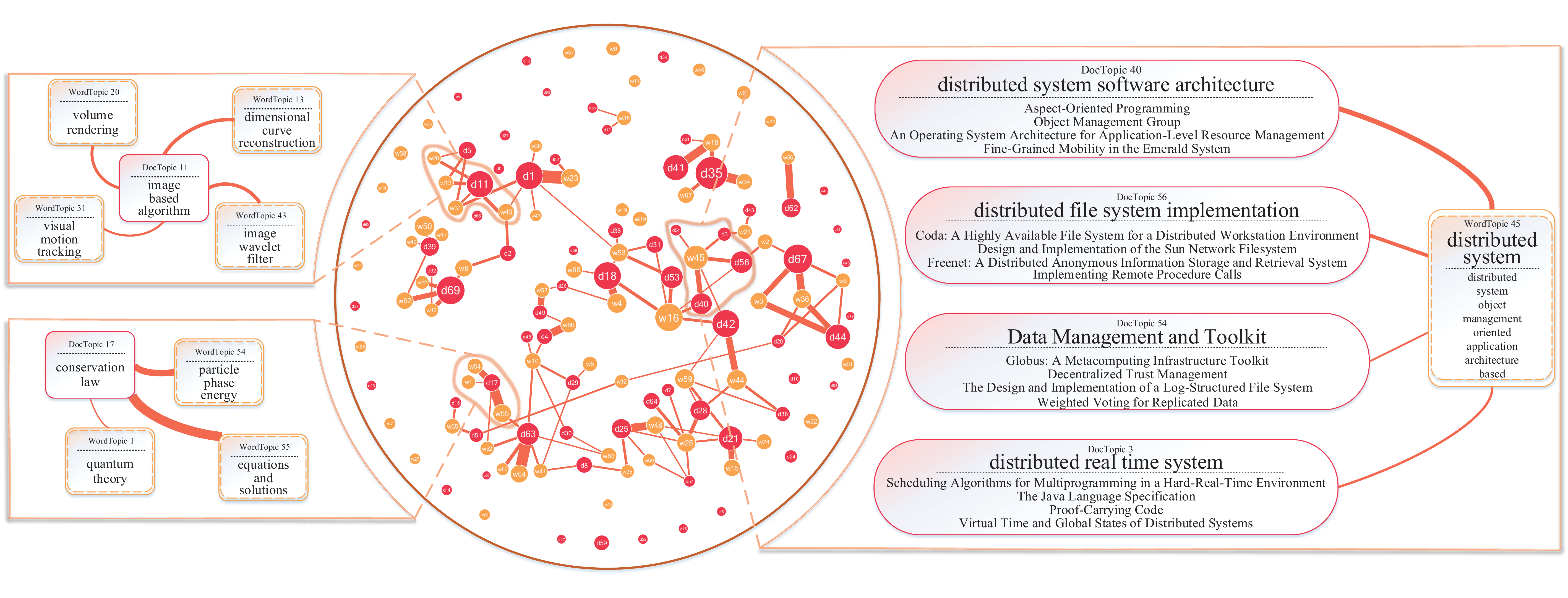}
\caption{Word-Doc Subgraph and some instances. The red nodes represent DocTopics and the orange nodes indicate WordTopics. Only the edges between WordTopics and DocTopics are displayed. Doctopic 11 and Doctopic 17 are expressed by indicative words.}
\label{relation}
\end{center}
\end{figure*}

\subsubsection{\textbf{Word-Doc Relation}} We summarize the contributions of Word-Doc relation from three perspectives. These examples are illustrated in  Figure~\ref{relation}.  

\emph{Connect WordTopic and DocTopic reasonably.} As Figure \ref{relation} suggests, the DocTopic d17 is about ``\emph{conservation law}'', and its neighbouring WordTopics are w54 ``\emph{particle phase energy}'', w1 ``\emph{quantum theory}'' and w55 ``\emph{equations and solutions}''. These topics cover some basic components of quantum mechanics. In addition, WordTopic w36 is about ``\emph{shared memory processor}'', and it has a strong link with DocTopic d44 ``\emph{shared memory system}'' and d67 ``\emph{cache performance}''. Also, it connects with DocTopic d20 ``\emph{power analysis of design}'' through a edge weighting about 15 since energy reduction plays an important role in shared memory processor. Besides, WordTopic w57 ``\emph{mobile robot navigation}'' is connected with DocTopic d49 ``\emph{mobile robot localization}'' and  d26 ``\emph{motion planning}''. These connections expose the main structure of ``\emph{mobile navigation}''. There are a lot of other examples in our heterogeneous topic web, readers can check them in our demo TopicAtlas.

\emph{Link WordTopics indirectly.} The missing co-occurrence phenomenon between WordTopics results in difficulty in spotting relevant WordTopics. However, DocTopics can serve as intermediaries  between WordTopics and uncover the hidden relationship. More specifically, if two WordTopics co-occur frequently with the same DocTopic, then we can say the two WordTopics are related. For example, WordTopic w43 ``\emph{image wavelet filter}'' is connected with WordTopics w13 ``\emph{dimensional curve reconstruction}'', w20 ``\emph{volume rendering}'' and w31 ``\emph{visual motion tracking}'' through  DocTopic d11 ``\emph{image based algorithm}'', which agrees with the fact that many volume rendering and visual motion tracking models are wavelet-based.



\emph{Locate Relevant Documents.} Through establishing connection between DocTopics and WordTopics, users can investigate relevant documents for WordTopics. Note that instead of simply recognizing all related documents for WordTopics, TopicAtlas organizes the relevant documents according to DocTopics and allows for inspecting them in different aspects . If a researcher aims to find relevant documents for WordTopic w45 ``\emph{distributed system}'', he can locate papers about the implementation of distributed file or network system in d56, examine distributed system architecture stuff in d40, get to know some data management or toolkit documents in distributed system from d54, or explore papers about distribution application in real-time system from d3. With the relevant documents sorted, the researcher is less prone to be swamped by the flood of information. 

\subsection{Topic Modeling}
\begin{figure*}[!t]
\centering
\subfloat[WordTopic on AAN]{\includegraphics[scale=0.2]{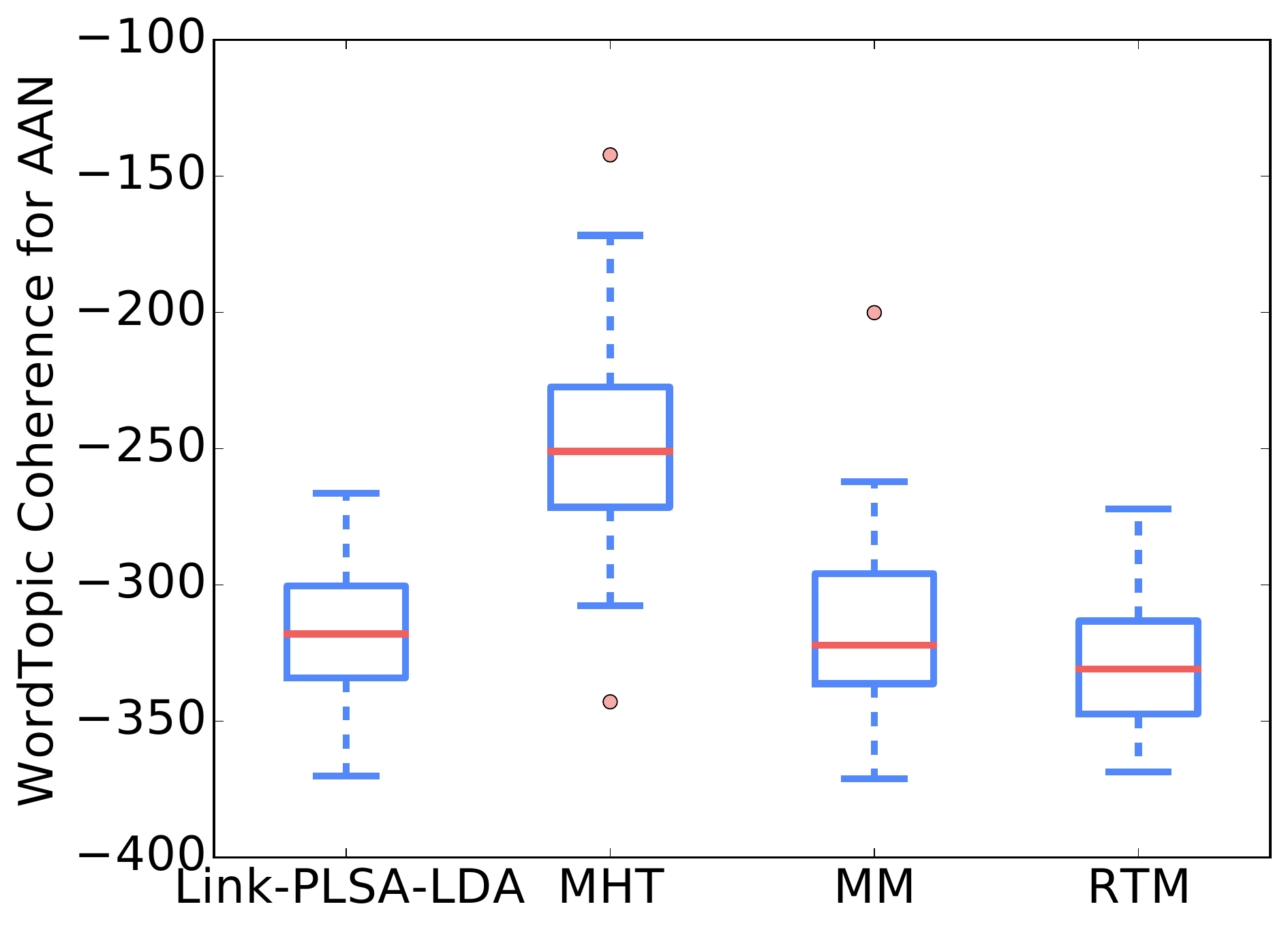}
\label{word_aan}}
\hfil
\subfloat[WordTopic on CiteseerX]{\includegraphics[scale=0.2]{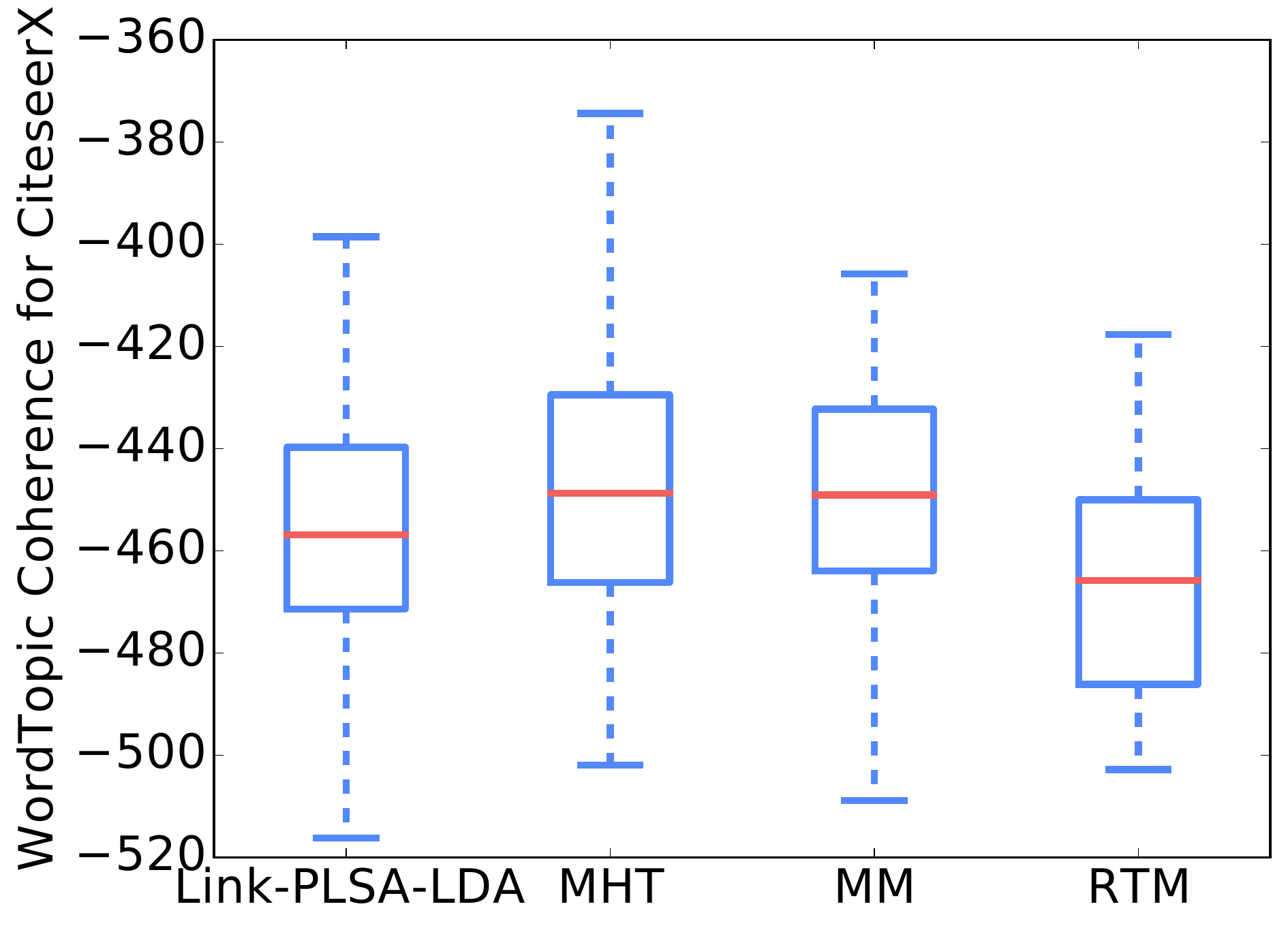}
\label{word_citeseer}}
\hfil
\subfloat[DocTopic on AAN]{\includegraphics[scale=0.2]{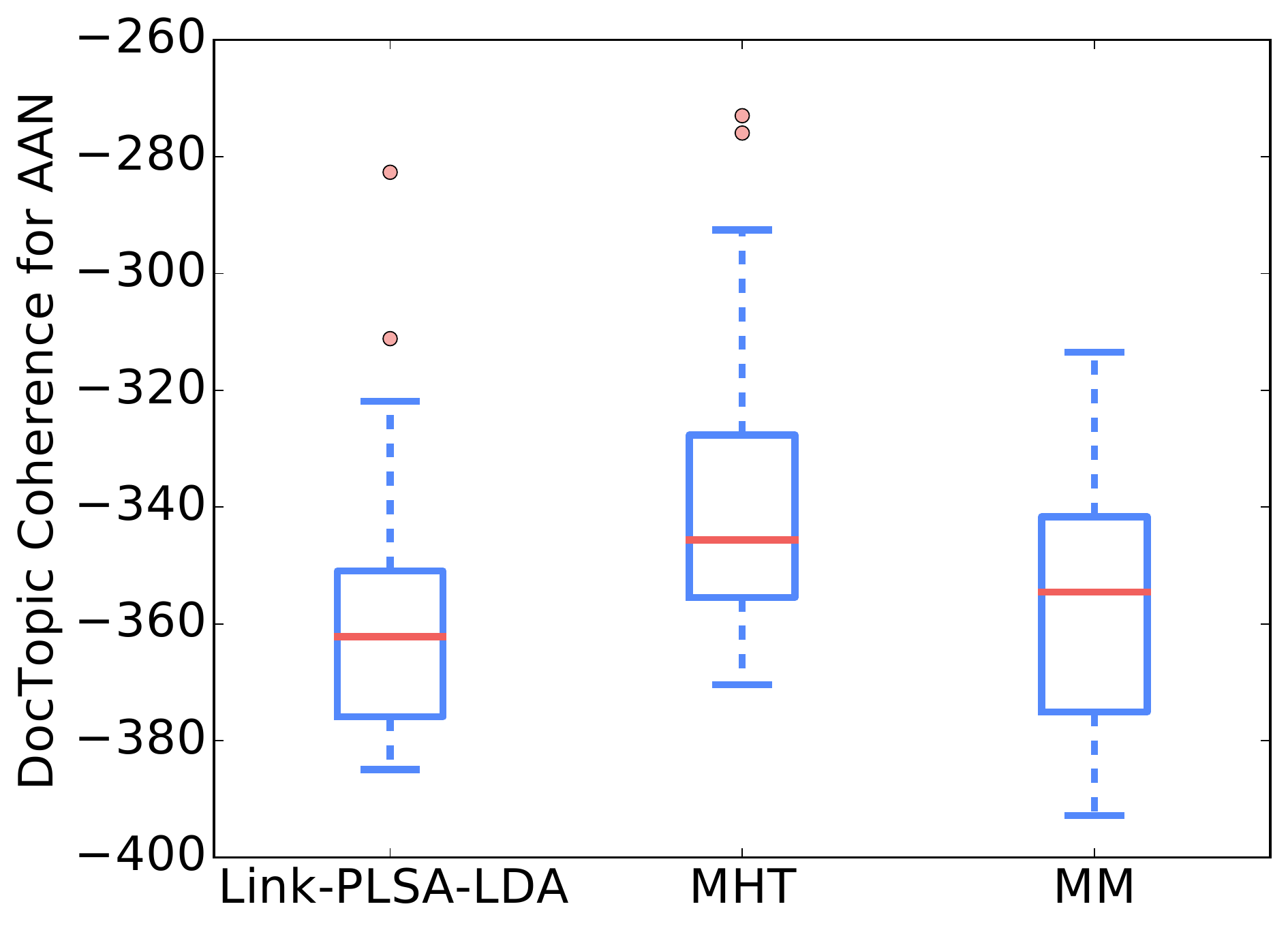}
\label{doc_aan}}
\hfil
\subfloat[DocTopic on CiteseerX]{\includegraphics[scale=0.2]{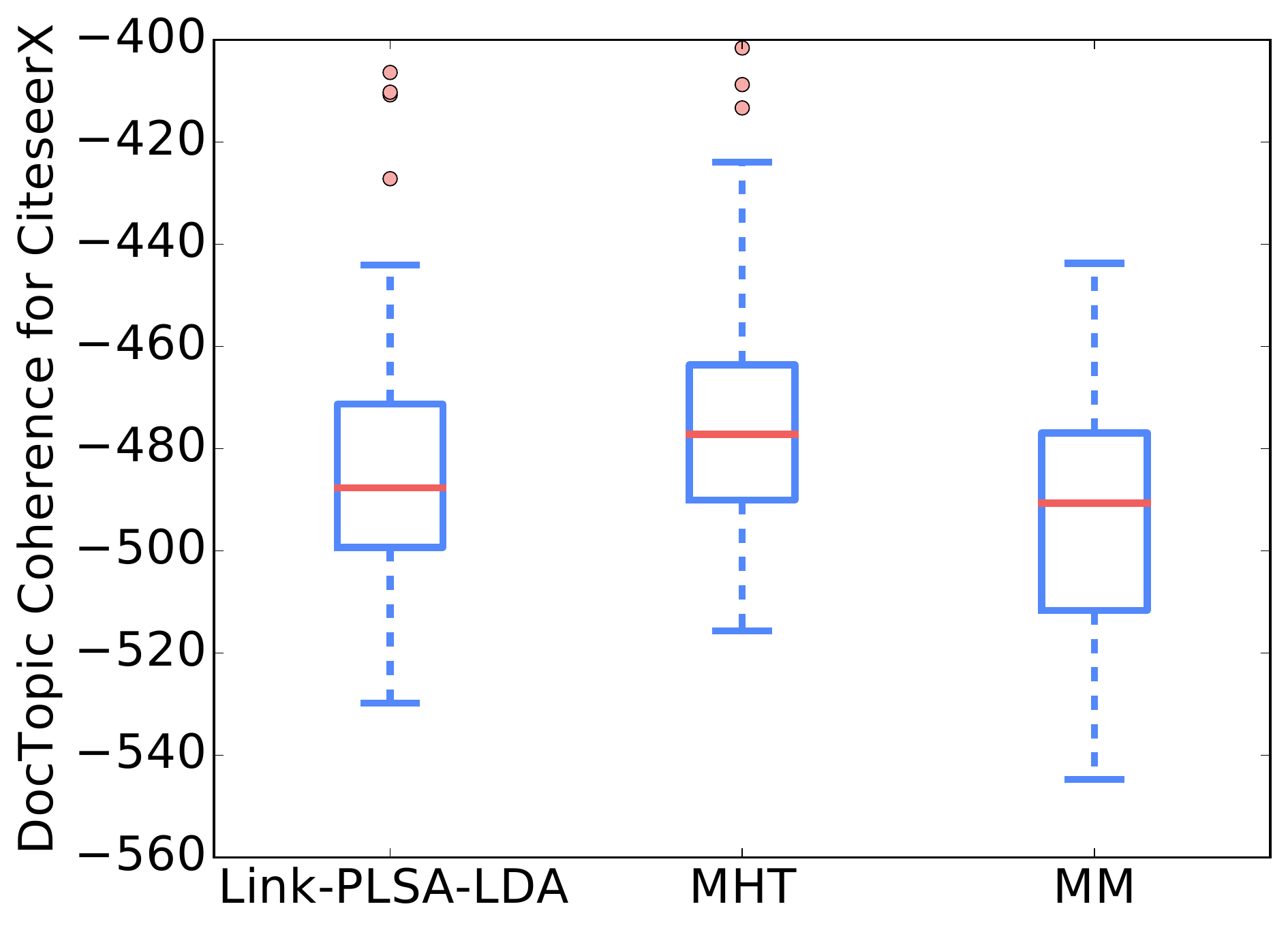}
\label{doc_citeseer}}
\caption{Topic coherence for WordTopic and DocTopic in two datasets (higher is better). }
\label{topic coherence}
\end{figure*}

Since we aim to obtain effective heterogeneous topic web, it is important to ensure that the introduction of the transition parameter has not come at the expense of the semantic quality of topics and the generalizability of the topic model.

\subsubsection{Comparative Methods} We compare our method MHT with mixed-membership model (MM) \cite{linklda}, Link-PLSA-LDA \cite{linkplsalda} and RTM \cite{chang2009relational}, all of which are joint models for both text and links. 
Mixed membership model is proposed by Erosheva et al. to classify documents~\cite{linklda}. Nallapati et al.~\cite{linkplsalda} propose two well-known joint topic models Pairwise-Link-LDA and Link-PLSA-LDA. Pairwise-Link-LDA models the presence and absence of links in a pairwise manner while Link-PLSA-LDA views links as ``link tokens''. Since Link-PLSA-LDA outperforms Pairwise-Link-LDA with respect to heldout likelihood and recall, we only include Link-PLSA-LDA in our baseline methods. The core idea of RTM is that topic relations directly account for the presence of links. 
To guarantee the justness, all these models are inferred through variational EM algorithm and parameters are initialized with the same way as MHT.

\subsubsection{Topic Interpretability} 
There are some metrics for evaluating topic interpretability such as \emph{PMI}~\cite{newman2010automatic}, \emph{word intrusion}~\cite{chang2009reading}, and \emph{topic coherence}~\cite{mimno2011optimizing}. We adopt \emph{topic coherence} in our experiment. For one thing, while word intrusion needs expert annotations, topic coherence is an automated evaluation metric and does not rely on human annotators. For another, topic coherence does not reference collections outside the training data as PMI dose. Also, topic coherence is proven more closely associated with the expert annotations than PMI~\cite{mimno2011optimizing}. 
Although it is originally designed for WordTopics, by using the indicative words as keywords, we can also calculate the topic coherence for DocTopics.
To distinguish the two different topic coherence score, we denote them as \emph{WordTopic coherence} and \emph{DocTopic coherence}.

We compare the topic coherence score of different methods for all topics, and the results are illustrated in  Figure~\ref{topic coherence}. As RTM does not produce DocTopics, it is not included in the comparison. Obviously, our model preserves comparable topic qualities to the baseline methods.

\subsubsection{Held-Out Log Likelihood}
Held-out Log Likelihood is a well-accepted metric to measure the generalizability and predictive power of topic models. To ease the favor for text and obtain a convincing result, we filter out the documents with less than 3 links and 8 links for AAN and CiteseerX respectively, and get a collection of AAN with $16,350$ documents and CiteseerX with $61,901$ documents.

 \begin{figure}[!htbp]
\centering
\subfloat[AAN]{\includegraphics[scale=0.23]{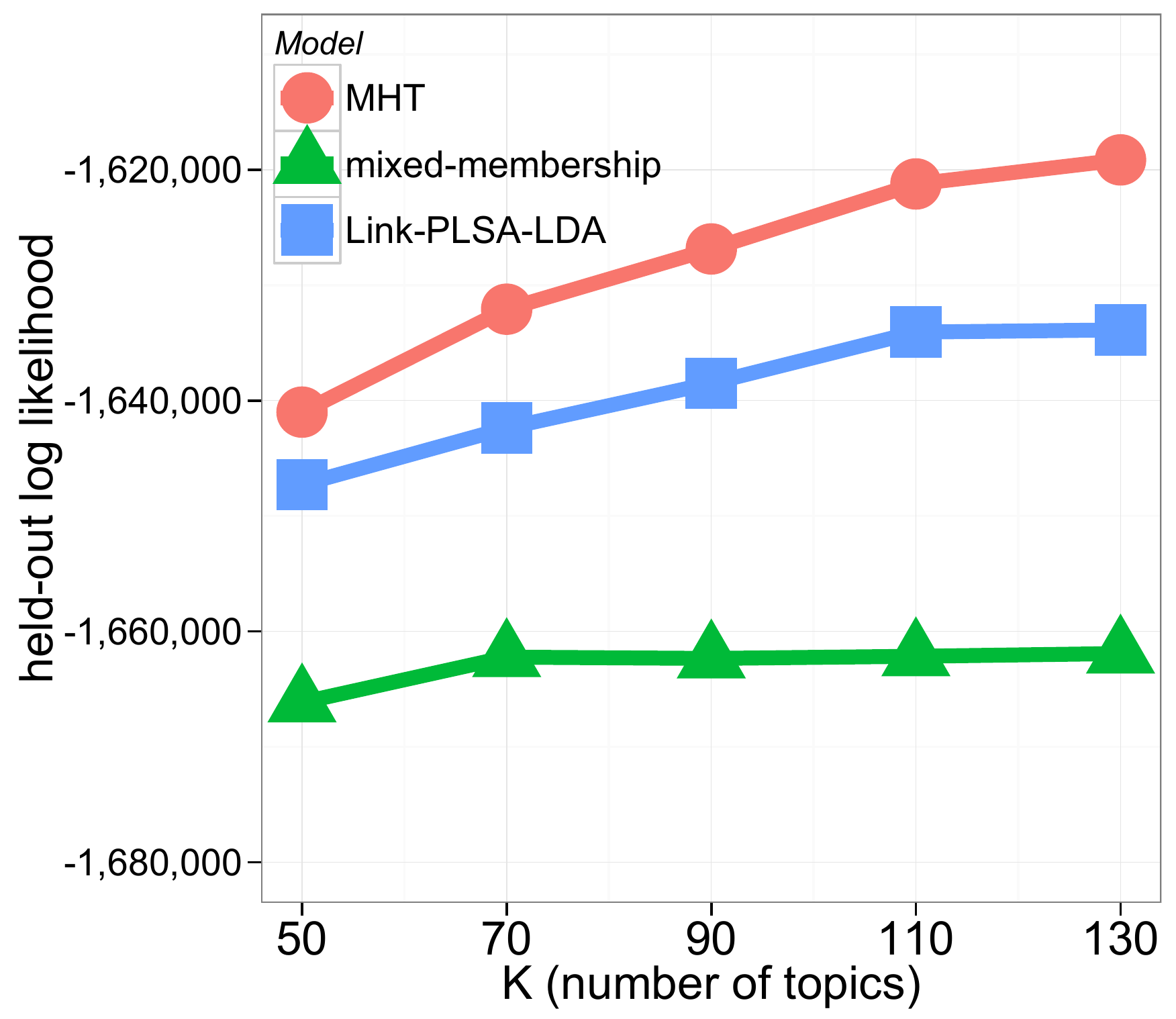}
\label{descrip1}}
\hfil
\subfloat[CiteseerX]{\includegraphics[scale=0.23]{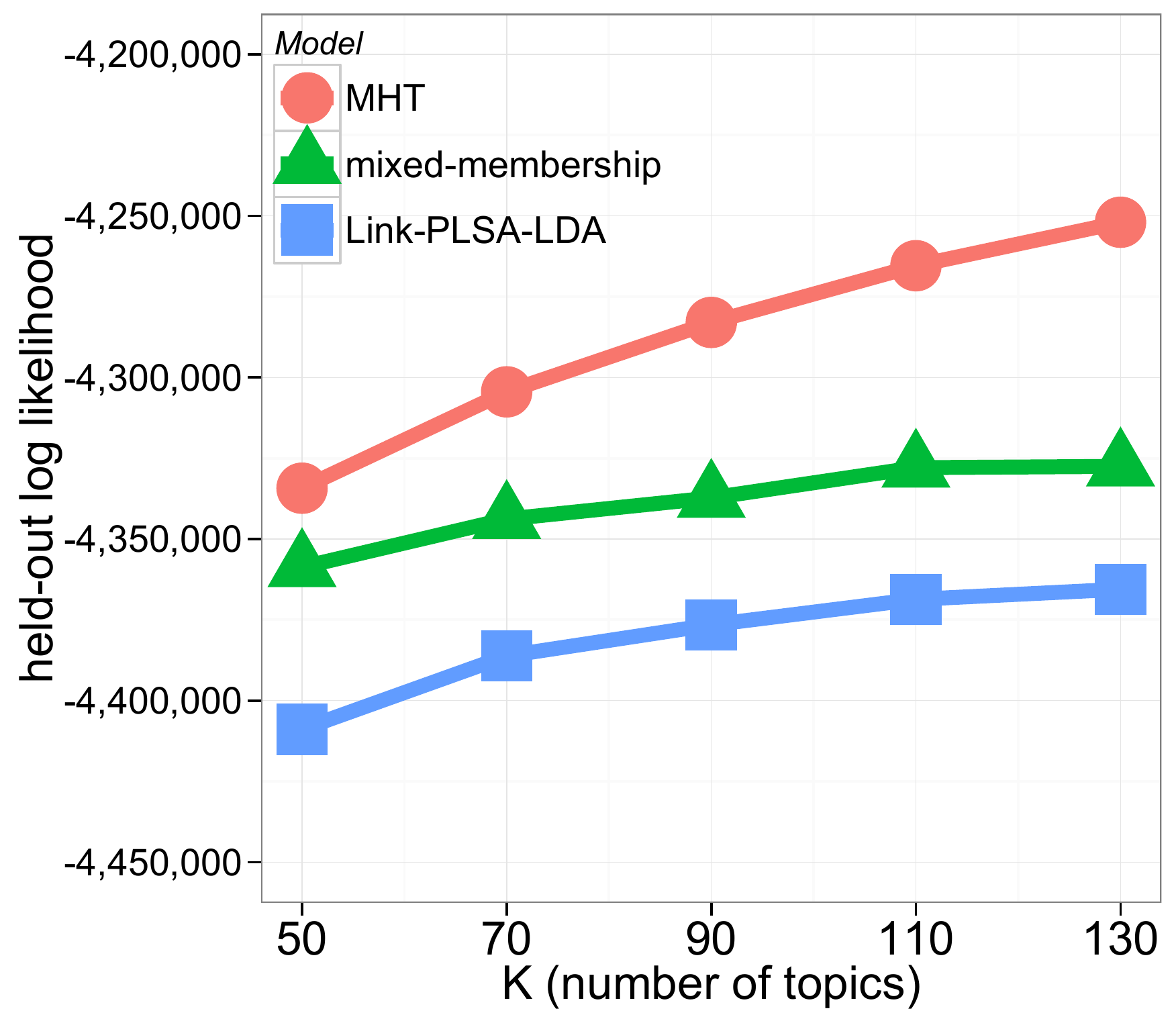}
\label{descrip2}}
\caption{Held-out log likelihood for both text and links on two datasets. (higher is better)} \label{heldout}
\end{figure}
Our experimental set-up is as follows. We randomly split data into five folds and repeat the experiment for five times, for each time we use one fold for test, four folds for training, and we report the average values in Figure \ref{heldout}. The performance of MHT is better than the baseline methods. Note that we exclude RTM in this part since held-out log likelihood favors RTM significantly due to its pairwise manner. 

\section{Conclusion}
\label{secConclusion}
In this paper, we present \emph{MHT}, short for \emph{Model for Heterogeneous Topic web}, a unified generative model involving two types of topics, namely WordTopic and DocTopic. The relationships between the two types of topics, Word-Word relation, Doc-Doc relation and Word-Doc relation, are quantified, based on which we construct the heterogeneous web of topics. In the experiment, we construct the heterogeneous topic web of AAN and CiteseerX collection and build a prototype demo system, called \emph{TopicAtlas} to exhibit the heterogeneous topic web and assist users' exploration. Qualitative analyses are presented to demonstrate the effectiveness of TopicAtlas. Besides, MHT shows good performance as a topic model with respect to topic interpretability and held-out log likelihood. 

\bibliographystyle{IEEEtran}

\begin{thebibliography}{10}
\providecommand{\url}[1]{#1}
\csname url@samestyle\endcsname
\providecommand{\newblock}{\relax}
\providecommand{\bibinfo}[2]{#2}
\providecommand{\BIBentrySTDinterwordspacing}{\spaceskip=0pt\relax}
\providecommand{\BIBentryALTinterwordstretchfactor}{4}
\providecommand{\BIBentryALTinterwordspacing}{\spaceskip=\fontdimen2\font plus
\BIBentryALTinterwordstretchfactor\fontdimen3\font minus
  \fontdimen4\font\relax}
\providecommand{\BIBforeignlanguage}[2]{{%
\expandafter\ifx\csname l@#1\endcsname\relax
\typeout{** WARNING: IEEEtran.bst: No hyphenation pattern has been}%
\typeout{** loaded for the language `#1'. Using the pattern for}%
\typeout{** the default language instead.}%
\else
\language=\csname l@#1\endcsname
\fi
#2}}
\providecommand{\BIBdecl}{\relax}
\BIBdecl

\bibitem{Marchionini2006}
\BIBentryALTinterwordspacing
G.~Marchionini, ``Exploratory search: From finding to understanding,''
  \emph{Commun. ACM}, vol.~49, no.~4, pp. 41--46, Apr. 2006. [Online].
  Available: \url{http://doi.acm.org/10.1145/1121949.1121979}
\BIBentrySTDinterwordspacing

\bibitem{klein2015exploratory}
L.~F. Klein, J.~Eisenstein, and I.~Sun, ``Exploratory thematic analysis for
  digitized archival collections,'' \emph{Digital Scholarship in the
  Humanities}, p. fqv052, 2015.

\bibitem{sinclair2003computer}
S.~Sinclair, ``Computer-assisted reading: Reconceiving text analysis,''
  \emph{Literary and Linguistic Computing}, vol.~18, no.~2, pp. 175--184, 2003.

\bibitem{gretarsson2012topicnets}
B.~Gretarsson, J.~O’donovan, S.~Bostandjiev, T.~H{\"o}llerer, A.~Asuncion,
  D.~Newman, and P.~Smyth, ``Topicnets: Visual analysis of large text corpora
  with topic modeling,'' \emph{ACM Transactions on Intelligent Systems and
  Technology (TIST)}, vol.~3, no.~2, p.~23, 2012.

\bibitem{alexander2014serendip}
E.~Alexander, J.~Kohlmann, R.~Valenza, M.~Witmore, and M.~Gleicher, ``Serendip:
  Topic model-driven visual exploration of text corpora,'' in \emph{Visual
  Analytics Science and Technology (VAST), 2014 IEEE Conference on}.\hskip 1em
  plus 0.5em minus 0.4em\relax IEEE, 2014, pp. 173--182.

\bibitem{linklda}
E.~Erosheva, S.~Fienberg, and J.~Lafferty, ``Mixed-membership models of
  scientific publications,'' \emph{Proceedings of the National Academy of
  Sciences}, vol. 101, no. suppl 1, pp. 5220--5227, 2004.

\bibitem{citationlda}
X.~Wang, C.~Zhai, and D.~Roth, ``Understanding evolution of research themes: a
  probabilistic generative model for citations,'' in \emph{Proceedings of the
  19th ACM SIGKDD international conference on Knowledge discovery and data
  mining}.\hskip 1em plus 0.5em minus 0.4em\relax ACM, 2013, pp. 1115--1123.

\bibitem{blei2006correlated}
D.~Blei and J.~Lafferty, ``Correlated topic models,'' \emph{Advances in neural
  information processing systems}, vol.~18, p. 147, 2006.

\bibitem{pairwise}
R.~M. Nallapati, A.~Ahmed, E.~P. Xing, and W.~W. Cohen, ``Joint latent topic
  models for text and citations,'' in \emph{Proceedings of the 14th ACM SIGKDD
  international conference on Knowledge discovery and data mining}.\hskip 1em
  plus 0.5em minus 0.4em\relax ACM, 2008, pp. 542--550.

\bibitem{chang2009relational}
J.~Chang and D.~M. Blei, ``Relational topic models for document networks,'' in
  \emph{International conference on artificial intelligence and statistics},
  2009, pp. 81--88.

\bibitem{ITM}
Q.~He, B.~Chen, J.~Pei, B.~Qiu, P.~Mitra, and L.~Giles, ``Detecting topic
  evolution in scientific literature: how can citations help?'' in
  \emph{Proceedings of the 18th ACM conference on Information and knowledge
  management}.\hskip 1em plus 0.5em minus 0.4em\relax ACM, 2009, pp. 957--966.

\bibitem{nallapati2011topicflow}
R.~Nallapati, D.~A. Mcfarland, and C.~D. Manning, ``Topicflow model:
  Unsupervised learning of topic-specific influences of hyperlinked
  documents.'' in \emph{AISTATS}, 2011, pp. 543--551.

\bibitem{weng2014topic}
L.~Weng and T.~M. Lento, ``Topic-based clusters in egocentric networks on
  facebook.'' in \emph{ICWSM}, 2014.

\bibitem{wang2015constructing}
C.~Wang, J.~Liu, N.~Desai, M.~Danilevsky, and J.~Han, ``Constructing topical
  hierarchies in heterogeneous information networks,'' \emph{Knowledge and
  Information Systems}, vol.~44, no.~3, pp. 529--558, 2015.

\bibitem{chaney2012visualizing}
A.~J.-B. Chaney and D.~M. Blei, ``Visualizing topic models.'' in \emph{ICWSM},
  2012.

\bibitem{maiya2014topic}
A.~S. Maiya and R.~M. Rolfe, ``Topic similarity networks: visual analytics for
  large document sets,'' in \emph{Big Data (Big Data), 2014 IEEE International
  Conference on}.\hskip 1em plus 0.5em minus 0.4em\relax IEEE, 2014, pp.
  364--372.

\bibitem{exploratory_jahnichen}
P.~Jahnichen, P.~Oesterling, G.~Heyer, T.~Liebmann, G.~Scheuermann, and
  C.~Kuras, ``Exploratory search through visual analysis of topic models,''
  \emph{Digital Humanities Quarterly (special issue)}, 2015.

\bibitem{plsa}
T.~Hofmann, ``Probabilistic latent semantic indexing,'' in \emph{Proceedings of
  the 22nd annual international ACM SIGIR conference on Research and
  development in information retrieval}.\hskip 1em plus 0.5em minus 0.4em\relax
  ACM, 1999, pp. 50--57.

\bibitem{blei2003latent}
D.~M. Blei, A.~Y. Ng, and M.~I. Jordan, ``Latent dirichlet allocation,''
  \emph{the Journal of machine Learning research}, vol.~3, pp. 993--1022, 2003.

\bibitem{ITM2}
L.~Dietz, S.~Bickel, and T.~Scheffer, ``Unsupervised prediction of citation
  influences,'' in \emph{Proceedings of the 24th international conference on
  Machine learning}.\hskip 1em plus 0.5em minus 0.4em\relax ACM, 2007, pp.
  233--240.

\bibitem{mei2008topic}
Q.~Mei, D.~Cai, D.~Zhang, and C.~Zhai, ``Topic modeling with network
  regularization,'' in \emph{Proceedings of the 17th international conference
  on World Wide Web}.\hskip 1em plus 0.5em minus 0.4em\relax ACM, 2008, pp.
  101--110.

\bibitem{linkplsalda}
R.~Nallapati and W.~W. Cohen, ``Link-plsa-lda: A new unsupervised model for
  topics and influence of blogs.'' in \emph{ICWSM}, 2008.

\bibitem{liu2009topic}
Y.~Liu, A.~Niculescu-Mizil, and W.~Gryc, ``Topic-link lda: joint models of
  topic and author community,'' in \emph{proceedings of the 26th annual
  international conference on machine learning}.\hskip 1em plus 0.5em minus
  0.4em\relax ACM, 2009, pp. 665--672.

\bibitem{le2014probabilistic}
T.~Le and H.~W. Lauw, ``Probabilistic latent document network embedding,'' in
  \emph{Data Mining (ICDM), 2014 IEEE International Conference on}.\hskip 1em
  plus 0.5em minus 0.4em\relax IEEE, 2014, pp. 270--279.

\bibitem{phits2}
D.~Cohn and H.~Chang, ``Learning to probabilistically identify authoritative
  documents,'' in \emph{ICML}.\hskip 1em plus 0.5em minus 0.4em\relax Citeseer,
  2000, pp. 167--174.

\bibitem{phits}
D.~Cohn and T.~Hofmann, ``The missing link-a probabilistic model of document
  content and hypertext connectivity,'' \emph{Advances in neural information
  processing systems}, pp. 430--436, 2001.

\bibitem{aan_source}
D.~R. Radev, P.~Muthukrishnan, and V.~Qazvinian, ``The acl anthology network
  corpus,'' in \emph{Proceedings of the 2009 Workshop on Text and Citation
  Analysis for Scholarly Digital Libraries}.\hskip 1em plus 0.5em minus
  0.4em\relax Association for Computational Linguistics, 2009, pp. 54--61.

\bibitem{mimno2011optimizing}
D.~Mimno, H.~M. Wallach, E.~Talley, M.~Leenders, and A.~McCallum, ``Optimizing
  semantic coherence in topic models,'' in \emph{Proceedings of the Conference
  on Empirical Methods in Natural Language Processing}.\hskip 1em plus 0.5em
  minus 0.4em\relax Association for Computational Linguistics, 2011, pp.
  262--272.

\bibitem{griffiths2004finding}
T.~L. Griffiths and M.~Steyvers, ``Finding scientific topics,''
  \emph{Proceedings of the National Academy of Sciences}, vol. 101, no. suppl
  1, pp. 5228--5235, 2004.

\bibitem{chang2009reading}
J.~Chang, S.~Gerrish, C.~Wang, J.~L. Boyd-Graber, and D.~M. Blei, ``Reading tea
  leaves: How humans interpret topic models,'' in \emph{Advances in neural
  information processing systems}, 2009, pp. 288--296.

\bibitem{newman2010automatic}
D.~Newman, J.~H. Lau, K.~Grieser, and T.~Baldwin, ``Automatic evaluation of
  topic coherence,'' in \emph{Human Language Technologies: The 2010 Annual
  Conference of the North American Chapter of the Association for Computational
  Linguistics}.\hskip 1em plus 0.5em minus 0.4em\relax Association for
  Computational Linguistics, 2010, pp. 100--108.

\end{thebibliography}

\end{document}